\PassOptionsToPackage{left=3.0cm, right=3.0cm, top=2.5cm, bottom=2.5cm}{geometry}
\documentclass[
  journal=jacsat,
  manuscript=article,
  layout=traditional,
]{achemso}

\usepackage[version=3]{mhchem} 
\usepackage{xcolor}
\usepackage{graphicx}
\usepackage{textcomp}
\DeclareUnicodeCharacter{2212}{\textminus}

\author{Tara P. Mishra}
\affiliation{Materials Sciences Division, Lawrence Berkeley National Laboratory, Berkeley, California 94720, United States}
\altaffiliation{Contributed equally}

\author{Zhuohan Li}
\affiliation{Materials Sciences Division, Lawrence Berkeley National Laboratory, Berkeley, California 94720, United States}
\altaffiliation{Contributed equally}

\author{Meghan Shen}
\affiliation{National Center for Electron Microscopy, Molecular Foundry, Lawrence Berkeley National Laboratory, Berkeley, California 94720, United States}
\alsoaffiliation{Department of Materials Science and Engineering, University of California, Berkeley, California 94720, United States}

\author{Maximilian Jaugstetter}
\affiliation{Chemical Sciences Division, Lawrence Berkeley National Laboratory, Berkeley, California 94720, United States}

\author{Livia P. Matte}
\affiliation{National Center for Electron Microscopy, Molecular Foundry, Lawrence Berkeley National Laboratory, Berkeley, California 94720, United States}
\alsoaffiliation{Department of Materials Science and Engineering, University of California, Berkeley, California 94720, United States}

\author{Jung O. Park}
\affiliation{Battery Material TU, Samsung Advanced Institute of Technology, Samsung Electronics, Co., Ltd.,Suwon, Gyeonggi-do, Republic of Korea}

\author{Hyunjin Kim}
\affiliation{Battery Material TU, Samsung Advanced Institute of Technology, Samsung Electronics, Co., Ltd.,Suwon, Gyeonggi-do, Republic of Korea}

\author{Benjamin Lam}
\affiliation{Department of Materials Science and Engineering, University of California, Berkeley, California 94720, United States}

\author{Karen Bustillo}
\affiliation{National Center for Electron Microscopy, Molecular Foundry, Lawrence Berkeley National Laboratory, Berkeley, California 94720, United States}

\author{Gerbrand Ceder}
\affiliation{Materials Sciences Division, Lawrence Berkeley National Laboratory, Berkeley, California 94720, United States}
\alsoaffiliation{Department of Materials Science and Engineering, University of California, Berkeley, California 94720, United States}
\email{gceder@berkeley.edu}

\author{Mary Scott}
\affiliation{Department of Materials Science and Engineering, University of California, Berkeley, California 94720, United States}
\alsoaffiliation{National Center for Electron Microscopy, Molecular Foundry, Lawrence Berkeley National Laboratory, Berkeley, California 94720, United States}
\email{mary.scott@berkeley.edu}

\title[An \textsf{achemso} demo]
  {
  Investigating the Degradation of LATP Solid Electrolyte in High Alkaline \ce{Li-O2} Batteries}

\abbreviations{IR,NMR,UV}

\begin{document}







\begin{abstract}

In this study, we address the challenge of electrolyte degradation in all-solid-state humidified \ce{Li-O2} batteries, which offer high theoretical energy density and potential cost advantages over conventional lithium-ion batteries. Combining STEM-EELS, and XPS characterizations with DFT calculations, we reveal the leaching of \ce{PO4^{3-}} and \ce{Al^{3+}} ions from the \ce{Li_{1.3}Al_{0.3}Ti_{1.7}(PO4)_3} (LATP) solid electrolyte upon battery discharge, caused by the highly alkaline environment. Upon charging, the leached ions precipitate as \ce{Li3PO4} and \ce{AlPO4}, which accumulate on the LATP surface and contribute to battery degradation. A Ti-rich layer is observed at the surface after a few cycles due to depletion of other cations. Our findings suggest that the degradation products are formed through repeated dissolution and precipitation in the discharge-charge cycles. Furthermore, our results indicate that the Ti-rich layer on the LATP surface can potentially reduce parasitic reactions. Our study provides mechanistic understanding of LATP solid electrolyte degradation in humidified \ce{Li-O2} cell, paving the way for designing more durable and efficient \ce{Li-O2} batteries.
\end{abstract}
\doublespacing
\section{Introduction}

The development of advanced energy storage systems has become paramount in addressing the growing demand for sustainable and efficient power sources. Among the various emerging technologies, lithium-air (\ce{Li-O2}) batteries have garnered significant attention \cite{lee2011metal,capasso2014experimental,au2022beyond}. The usage of \ce{O2} from the air as the active material allows a Li-\ce{O2} battery to achieve much higher gravimetric ($\approx$3500 Wh/kg) and volumetric ($\approx$3400 Wh/L) energy density (the values are based on the fully discharged state) than conventional Li-ion batteries ($\approx$100-300 Wh/kg and $\approx$400-800 Wh/L), as well as potentially much lower cost as Li-\ce{O2} cathodes do not use expensive transition metals (e.g., Ni and Co). \cite{zheng2008theoretical,tan2013prediction,Ma2015_review,luntz2014nonaqueous,Tian2021_review}. In a non-aqueous environment, the discharge and charge of the \ce{Li-O2} are based on the formation of \ce{Li2O2}\cite{lu2010influence}:

\begin{equation}
\ce{2Li+} + 2\ce{e-} + \ce{O2} \ce{<=>} \ce{Li2O2}  \quad (\ce{E^0} = 2.96 \text{ V} \text{ vs. } \ce{Li}/\ce{Li+})
\end{equation}

Despite its high theoretical energy density, the conventional \ce{Li} - \ce{O2} battery faces some significant limitations. First, undesirable side reactions can occur between the highly oxidative radicals formed at the cathode during the charge and discharge cycles and organic liquid electrolytes \cite{McCloskey2011_solvent,freunberger2011reactions} or carbon cathodes \cite{mccloskey2012twin,ottakam2013carbon}. Secondly, the side reaction results in the formation of carbonates, which significantly increases the charge overpotential.\cite{mccloskey2012twin}. To overcome these issues, the use of all-solid-state \ce{Li}-\ce{O2} batteries has been proposed\cite{hong1978crystal,adachi1996fast,macfarlane1999lithium,knauth2009inorganic,chi2021highly,le2019highly,kim2022carbon}, where stable solid electrolytes and carbon-free cathodes are utilized. 

However in solid-state \ce{Li-O2} batteries, the insulating nature of the solid discharge product \ce{Li2O2} can lead to limited practical discharge capacity \cite{Radin2013_charge, luntz2013tunneling,radin2012electronic,viswanathan2011electrical,gerbig2013electron,radin2015dopants}. The limited electron transport restricts the continuous growth of \ce{Li2O2} once the initial discharge products passivate the original cathode surfaces, leading to the ``sudden death" of the Li-\ce{O2} cell during discharge \cite{zhang2022reacquainting}. To address this challenge, humidified \ce{O2} has been proposed, which chemically converts the growth-limited \ce{Li2O2} into soluble and higher conductivity \ce{LiOH} \cite{liu2015cycling,ma2020mixed,kim2022carbon}. In a humidified environment, the cell reaction also changes into a 4\ce{e-} process and follows the equation \cite{liu2015cycling,lu2014aprotic}:

\begin{equation}
\ce{4Li+} + 4\ce{e-} + \ce{O2} + 2\ce{H2O} \ce{<=>} 4\ce{LiOH} \quad (\ce{E^0} = 3.4 \text{ V} \text{ vs. } \ce{Li}/\ce{Li+})
\end{equation}

Besides a slight reduction in the specific capacity, humidified \ce{Li-O2} batteries have two significant advantages over traditional Li-\ce{O2} systems. Firstly, the humidity facilitates ion transport by converting the insoluble solid \ce{Li2O2} to soluble \ce{LiOH}, thereby increasing the discharge capacity \cite{gao2023recent,dai2019fundamental}. Secondly, the theoretical discharge voltage is increased from 2.96 to 3.4 V (vs \ce{Li/Li+}) thereby increasing the power output of the battery \cite{kim2022carbon}.

In light of this, several chemical platforms for \ce{Li-O2} batteries have been investigated, including LISICON, NASICON-type Li-ion conductors such as \ce{Li_{1+x}Al_xTi_{2-x}(PO4)3} (LATP) and \ce{Li_{1+x}Al_xGe_{2-x}(PO4)3} (LAGP), thio-LISICON, glass-ceramics, and solid organic electrolytes \cite{hong1978crystal,kumar2010cathodes,kichambare2012mesoporous,kanno2001lithium,kobayashi2008interfacial,mizuno2005new,kondo1992new,takada1993electrochemical,fergus2010ceramic}. Oxide Li-ion conductors such as garnets and layered oxides are sensitive to humidity and undergo \ce{Li+}/\ce{H+} exchange and surface degradation under ambient conditions \cite{Cheng2018_garnet_LHX, Wang2024_garnet_LHX,Zhou2024_garnet_LHX,zuo2022guidelines}, requiring additional surface treatments to maintain low surface impedance. In comparison, NASICON's offer better chemical stability in ambient atmosphere, has made it of significant interest \cite{chen2019approaching}. 

In humidified solid-state Li-\ce{O2} batteries a new challenge emerges  due to the formation of LiOH as the discharge product. This creates a highly alkaline condition that adversely affects the stability and ionic conductivity of NASICONs. \cite{Shimonishi2011_NASICON_alkaline, Zhang2013_NASICON_water,He2014_LAGP_water,Safanama2017_LAGP_water}. Previous reports have shown that during the operation of a humidified \ce{Li-O2} battery, large water droplets grow and shrink during discharge and charge, respectively \cite{kim2023operando}. These droplets result from the hydration of \ce{LiOH}, creating saturated \ce{LiOH} solutions with high pH (theoretically up to $\approx$15 in the limit of a saturated LiOH solution \cite{monnin2005thermodynamics}). The extremely high alkaline environment is corrosive to the solid electrolyte, leading to degradation and failure of the \ce{Li-O2} cells \cite{kim2022carbon}.

Previous studies have investigated the corrosion of NASICON electrolytes in aqueous solution environments \cite{hasegawa2009study,dermenci2020stability,Lam2024_ltgp_degradation}. Analysis of the dissolution products of \ce{LiGe_xTi_{2-x}(PO4)3} pellets and powders in highly alkaline solutions has demonstrated that increasing the \ce{Ti} content within the NASICON framework enhances the alkaline stability of these electrolytes \cite{Lam2024_ltgp_degradation}. In another study, Hasegawa \textit{et al.} investigated the effect of pH on \ce{Li_{1+x}Al_xTi_{2-x}(PO4)3} (LATP) solid electrolytes by immersing them in various solutions. They found that while LATP remains stable under neutral pH conditions, it degrades under both acidic and alkaline conditions \cite{hasegawa2009study}. These studies advance our understandings about how NASICON electrolytes behave in corrosive environments, especially at highly alkaline conditions. In the real battery settings, the NASICON electrolytes in the realistic battery settings are exposed to much more complex electrochemical conditions: while the pH at the electrolyte's surface becomes very high at the end of the discharge, the pH decreases when the discharge product \ce{LiOH} decomposes upon charging. To date, comprehensive characterization of electrolyte degradation under realistic cycling conditions remains limited. Such detailed characterization is essential for developing new strategies for all-solid-state \ce{Li}-\ce{O2} batteries. Additionally, establishing advanced characterization workflows for solid electrolytes under cycling conditions is crucial for understanding their chemical stability in corrosive environments across a wide range of electrochemistry applications \cite{hou2021multiscale,abdullahi2024nasicon}.

In this study, we conducted advanced characterization of a high-energy-density humidified all-solid-state \ce{Li}-\ce{O2} cell containing a planar Pt-patterned cathode on a \ce{LATP} solid electrolyte at different cycling stages using X-ray Photoelectron Spectroscopy (XPS) and Electron Energy Loss Spectroscopy (EELS). With the help of density functional theory (DFT), we rationalize the formation of the discharge products and the degradation process of the LATP upon cycling. Our results indicate that the degradation of LATP begins with the dissolution of \ce{PO4^{3-}} groups. The depletion of \ce{PO4^{3-}} groups on the LATP surface subsequently promotes the leaching of \ce{Al^{3+}} ions, while \ce{Ti^{4+}} ions are left behind on the top layer of LATP surface. The remaining \ce{Ti^{4+}} gradually accumulates and forms a Ti-rich layer on the surface, which acts as a protecting layer to reduce the further degradation of the LATP solid electrolyte. This study elucidates the microscopic evolution of phosphate NASICONs used as electrolytes in solid-state humidified \ce{Li-O2} batteries upon cycling and proposes strategies to mitigate parasitic degradation.

\section{Results}
\subsection{Surface investigation of LATP electrolyte upon cycling}

\begin{figure}
  \includegraphics[width=0.8\textwidth]{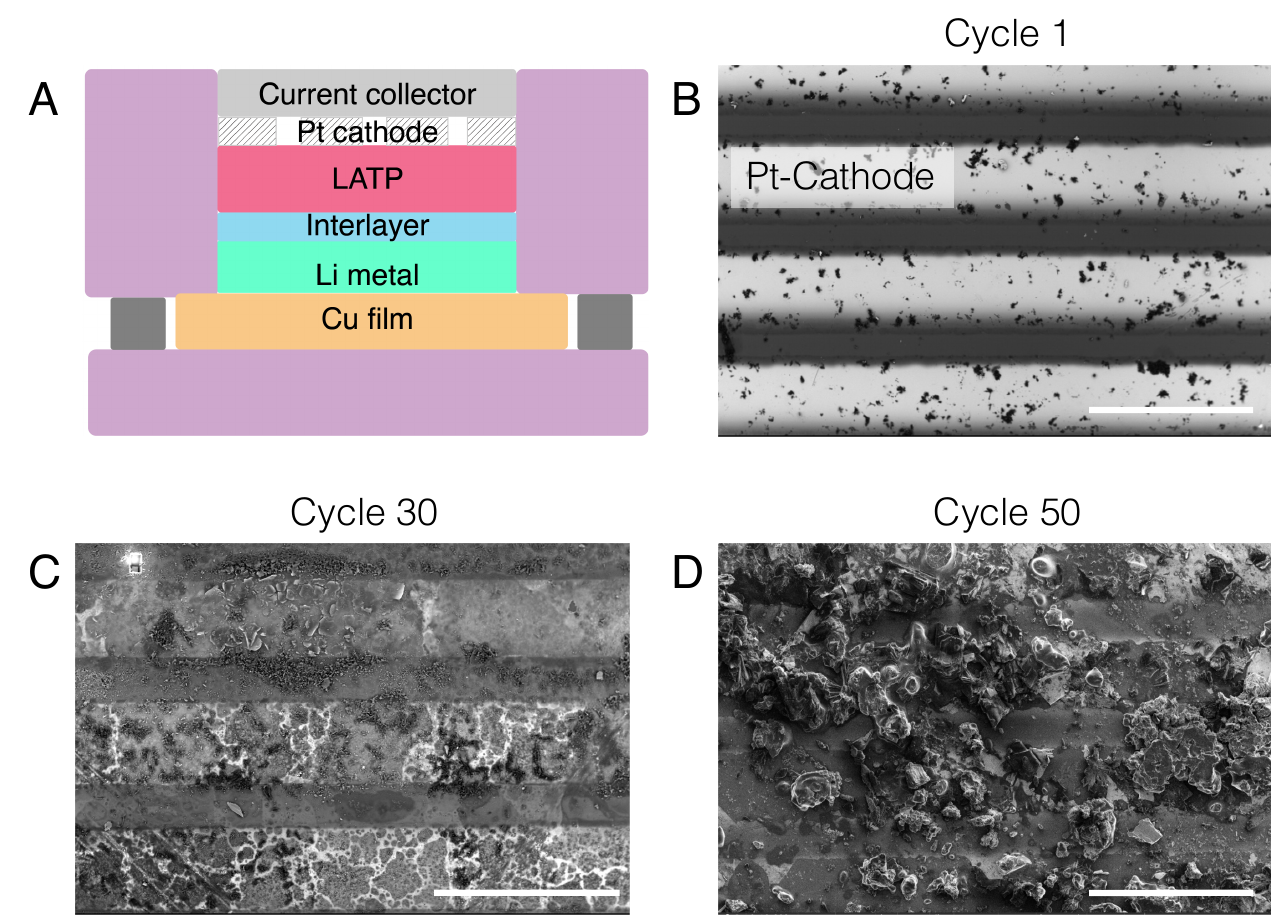}
  \caption{(A) Schematic illustration of the humidified \ce{Li-O2} battery configuration utilized for advanced characterization of the LATP solid electrolyte during cycling of the solid-state \ce{Li-O2} battery. SEM micrographs of the LATP solid electrolyte surface viewed top down, with patterned Pt cathodes (bright regions) after (B) 1 cycle, (C) 30 cycles, and (D) 50 cycles. Scale bars in all SEM images correspond to 300 $\mu$m.}
  \label{fgr:sem_surf}
\end{figure}

In this work, we aim to explore the effect of electrochemical cycling on the degradation of the NASICON-type solid electrolyte LATP (\ce{Li_{1.3}Al_{0.3}Ti_{1.7}(PO_4)_3}) in fully humidified all–solid-state lithium-air (\ce{Li-O2}) batteries. To track the evolution of the electrolyte, \ce{Li-O2} cells were galvanostatically cycled between 3.0 and 4.5 V vs. \ce{Li}/\ce{Li+}. A schematic cross-section of the cell architecture is provided in Figure~\ref{fgr:sem_surf}(A). Multiple cells were cycled for different number of the cycles (1, 30, and 50 cycles) to study the degradation of LATP solid electrolyte post cycling in the \ce{Li-O2} battery. The details of the battery setup and cycling procedures of humidified \ce{Li-O2} are described in the method sections. Voltage–capacity profiles for cells after the 1st and 30th cycles during both charge and discharge are shown in Supplementary Figures S1 and S2, respectively. The coulombic efficiency as a function of cycle number is shown in Supplementary Figures S3 and S4 for the \ce{Li-O2} battery cycled for 30 and 50 cycles, respectively. From these coulombic efficiency curves, we observe that the efficiency increases from below 80\% in the initial cycles ($\approx$ 5 to 10 cycles) to nearly 100\% after a few cycles, after which it remains constant. The \ce{Li-O2} cell cycled 50 times fails at the 49th cycle, as indicated by a sudden drop in Coulombic efficiency (Supplementary Figure S4). These electrochemistry results are similar to what has been observed before in similar \ce{Li-O2} batteries \cite{kim2023operando,kim2022carbon,ma2020mixed}.

At different cycle numbers (1, 30, and 50), the LATP solid electrolytes were extracted from the cell for further characterization. Figure~\ref{fgr:sem_surf} (B-D) presents low-magnification scanning electron microscopy (SEM) micrographs of the LATP electrolyte after varying numbers of cycles. After the first cycle, the Pt cathode (bright stripes in SEM images) is clearly visible on the LATP electrolyte (dark stripes), as shown in Fig.\ref{fgr:sem_surf} (B). After 30 cycles, some degradation products accumulate at the top of both the LATP electrolyte and Pt cathode, as seen in Fig.\ref{fgr:sem_surf} (C). Upon further cycling (50 cycles), significant corrosion is observed, and the Pt cathode is no longer discernible (Fig.~\ref{fgr:sem_surf} (D)). The Pt cathode is most likely buried under the discharge product (discussed in more detail below). From the evolution of voltage profiles, it is observed that the cell has negligible coulombic efficiency after 49 cycles (Supplementary Figure S4) and fails by the 50th cycle. We speculate that the cell failure results from the severe surface degradation as observed in Figure \ref{fgr:sem_surf} (D).

\begin{figure}
  \includegraphics[width=0.9\textwidth]{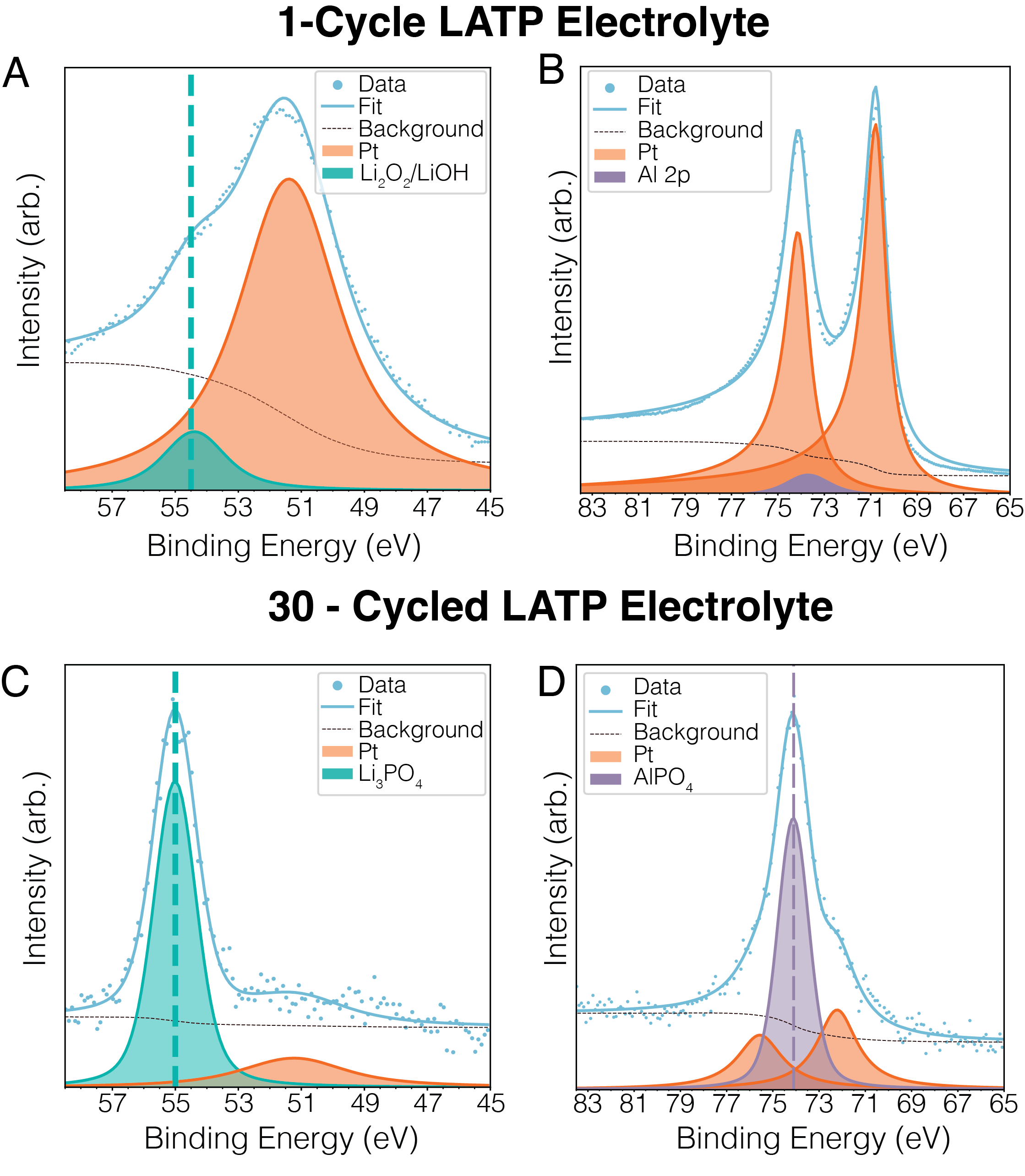}
  \caption{XPS spectra of the LATP solid electrolyte surface from the humidified \ce{Li-O2} battery, revealing degradation products after (A) 1 cycle, with Li 1s region showing LiOH formation, and (B) Al 2p region, and after (C) 30 cycles, with Li 1s region showing \ce{Li3PO4} formation, and (D) Al 2p region showing \ce{AlPO4} formation.}
  \label{fgr:xps_chara}
\end{figure}

To better understand the degradation process on the surface of the LATP pellets post-cycling, we undertake x-ray photoemission spectroscopy (XPS) on the LATP pellets to compare chemical composition of the degradation products after 1 and 30 cycles. The XPS peaks that we investigated were the Li-1s, Pt-5p, Pt-4f, Al-2p, C-1s, and O-1s peaks. After 1 cycle, the Pt-5p and Pt-4f peaks can be clearly observed in panels A and B of Fig.\ref{fgr:xps_chara}, respectively \cite{fryer2016synthesis,choi2024reevaluation}, consistent with the intact Pt cathode stripes shown in the SEM image (Figure \ref{fgr:sem_surf} (B)). After deconvolution of the Pt-5p peaks and Li-1s peak, a  Li peak at 54.5 eV is obtained, which corresponds to \ce{Li2O2} and/or \ce{LiOH} \cite{yao2013thermal} (Fig.\ref{fgr:xps_chara} (A)). These two components cannot be differentiated due to the small difference in their binding energy ( $\approx$ 0.1 eV).  Furthermore, a very small Al-2p peak at 74.5 eV is observed in the Al-2p region (Fig.~\ref{fgr:xps_chara}(B)). As no clear degradation is observed in the SEM image after the first cycle, we attribute this small Al-2p peak to the underlying LATP below the Pt cathode \cite{lindblad1994characterization}. The detailed deconvolution and the fitting procedure of the XPS spectra are described in the method section.

XPS measurements from LATP pellets after 30 cycles show a significant increase in the Li-1s peak as compared to the Pt-5p peaks(Fig.~\ref{fgr:xps_chara}(C)). The \ce{Li} 1s peak ($\approx$ 55 eV) after 30 cycles shows a shift to a higher binding energy compared to the \ce{Li}-1s peak in the 1 cycle case. This Li 1s peak position is in close agreement with that of \ce{Li3PO4}\cite{morgan1973inner}. Furthermore, in the region where the Al-2p peaks are present (Fig.~\ref{fgr:xps_chara}(D)), the Pt signal is significantly reduced when compared to the same energy region after 1 cycle (Fig.~\ref{fgr:xps_chara}(B)). This reduction is consistent with the reduction of the Pt-5p signal observed in the panel C of Fig.~\ref{fgr:xps_chara}. In the Al binding energy range, the Al-2p peak is clearly noticeable with a binding energy of 74.1 eV, in close agreement with that of \ce{AlPO4} (Fig.~\ref{fgr:xps_chara}(D))\cite{lindblad1994characterization}. These results suggest that the LATP solid electrolyte undergoes surface corrosion upon cycling in the humidified \ce{Li-O2} cell and that degradation products such as \ce{Li3PO4} and \ce{AlPO4} accumulate on the surface of the system. The reduced Pt signals in XPS measurements results from these degradation products covering the Pt cathode. This likely disrupts the electronic conduction pathway and ultimately leads to cell failure after prolonged cycling (Supplementary Figure S4).

\subsection{Cross-sectional characterization of the degradation of the LATP solid electrolyte}

\begin{figure}
  \includegraphics[width=0.9\textwidth]{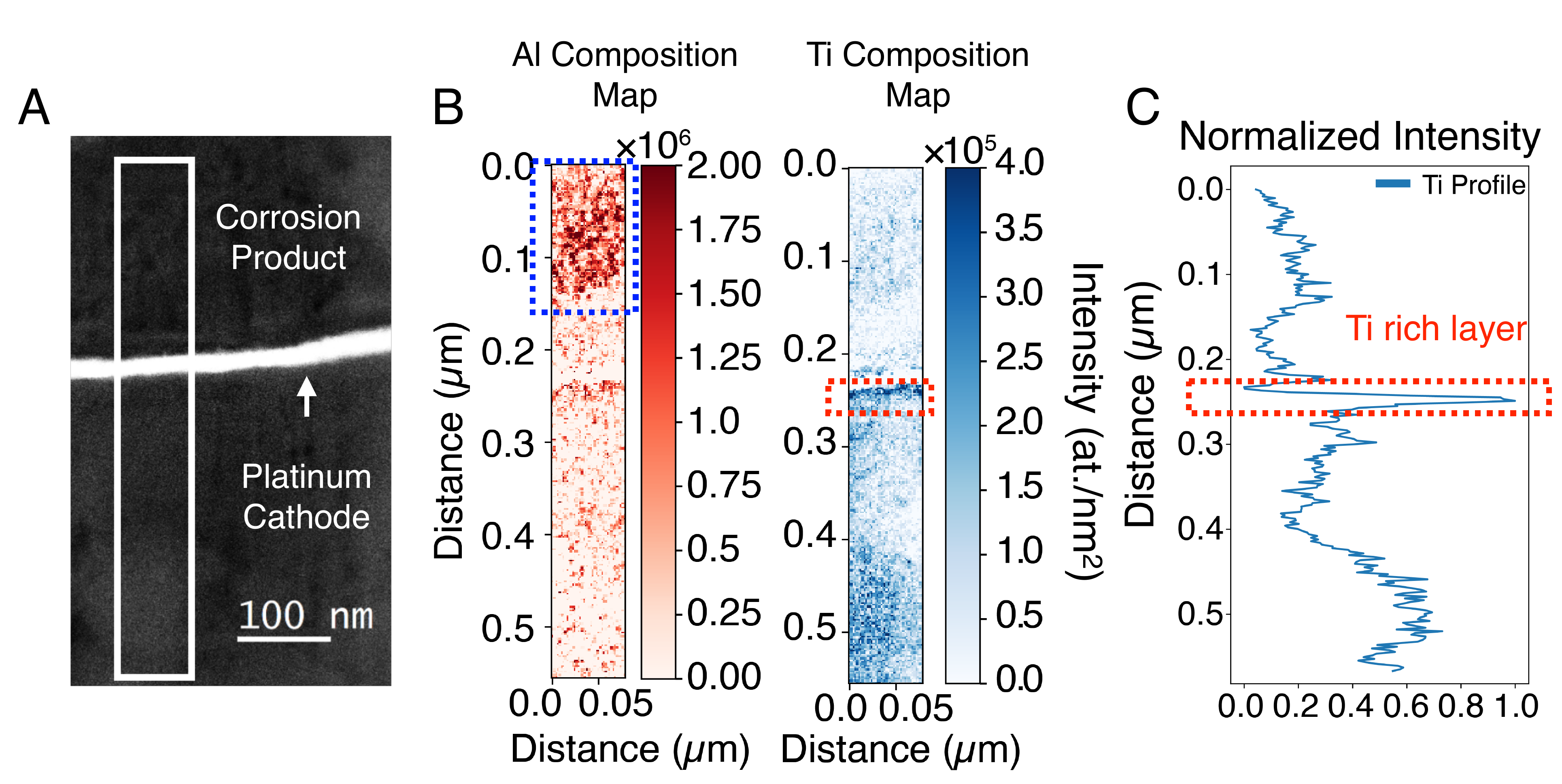}
  \caption{(A) Low-magnification STEM-HAADF micrograph showing a cross-sectional view of the LATP solid electrolyte after 30 cycles, highlighting the corrosion products and the Pt cathode region. (B) Al and Ti compositional maps obtained from EELS measurements taken within the region indicated by the white box in panel (A). The blue and red dashed boxes highlight areas with elevated Al and Ti concentrations, respectively. (C) Cross-sectionally integrated Ti composition profile, with the Ti-rich layer beneath the Pt cathode indicated by the red box. }
  \label{fgr:comp_mapping}
\end{figure}

To further characterize potential degradation of the LATP solid electrolyte beneath the surface, we performed a STEM-EELS measurements of a cross-sectional sample taken from near the surface of the LATP electrolyte after 30 cycles. A SEM image of the region of the FIB sample preparation is shown in supplementary figure S5. The STEM-HAADF image obtained from the FIB lamella is shown in Fig.~\ref{fgr:comp_mapping}(A). The Pt cathode can be seen in this cross-sectional image as a bright line. Consistent with the SEM (Fig.\ref{fgr:sem_surf}) and XPS (Fig. \ref{fgr:xps_chara}) measurements of the LATP pellet surface, the aggregation of the corrosion products is visible above the Pt cathode (the white line), which are initially absent in the LATP electrolyte post 1 cycle, as shown in supplementary figure S6.

The chemical composition of the degradation product is further investigated using EELS compositional mapping shown in Fig.~\ref{fgr:comp_mapping}. We observe that the corrosion products above the Pt cathode are significantly enriched in Al (Fig.~\ref{fgr:comp_mapping}(B)). However, there is little Ti in these post cycling corrosion products formed on the surface of the LATP electrolyte as evidenced by the Ti composition map. Interestingly, we note that right below Pt cathode a very thin ($\approx$ 20 nm) region of enriched Ti layer is formed post cycling (Fig.~\ref{fgr:comp_mapping}(B)). This can be clearly visualized from the laterally integrated Ti composition profile shown in Fig.~\ref{fgr:comp_mapping}(C). The enriched Ti layer formed during the cycling is marked with a red dashed rectangle. It is important to note here that such enriched Ti layer was not present in the cross-section of the post 1 cycled LATP sample as shown in supplementary figure S5. 

\begin{figure}
  \includegraphics[width=\textwidth]{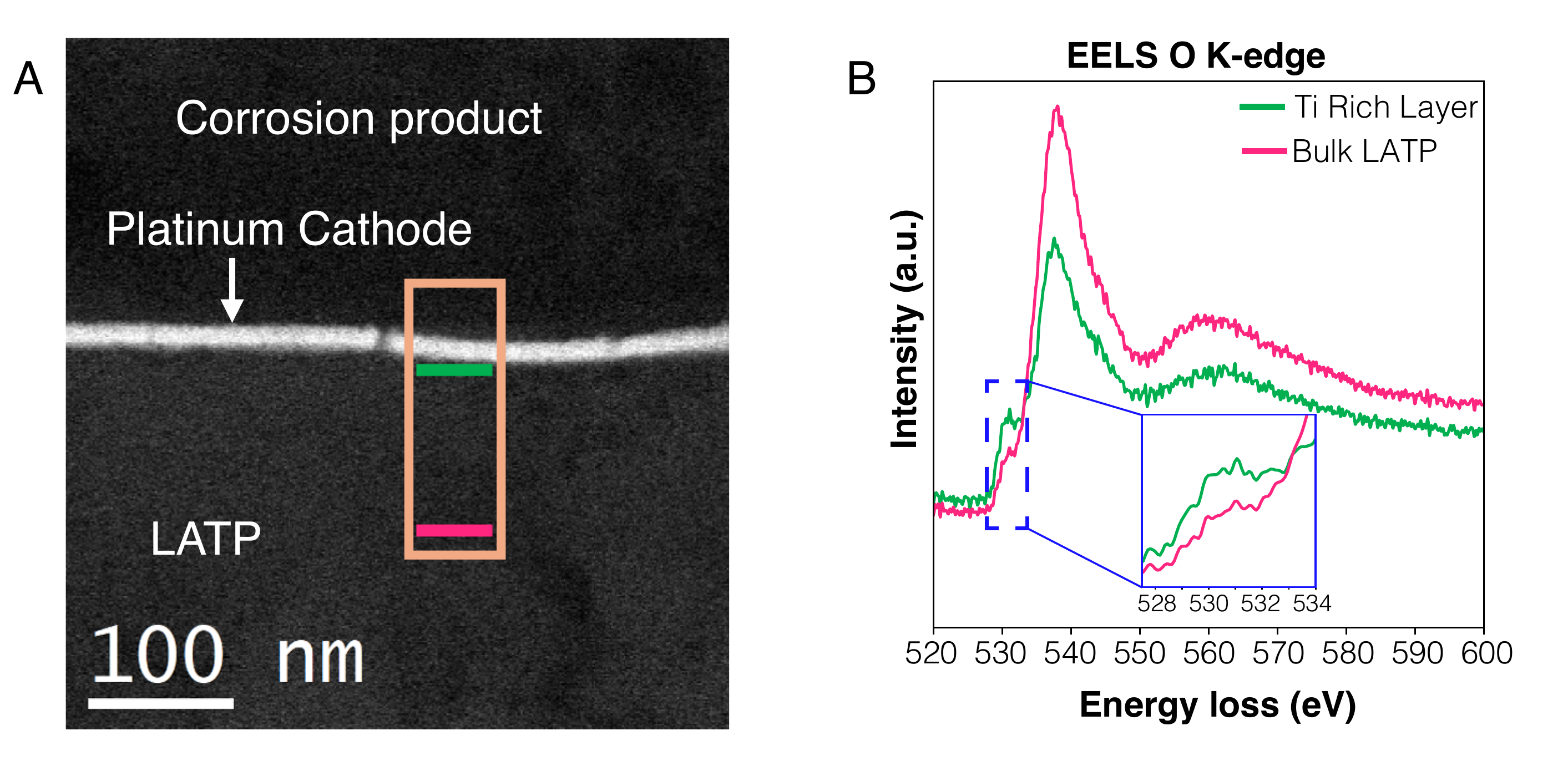}
  \caption{(A) Low-magnification cross-sectional image of the LATP solid electrolyte showing the Pt cathode, LATP solid electrolyte, and the corrosion product formed during the cycling of the humidified \ce{Li-O2} cell. (B) Oxygen K-edge obtained using EELS measurement from the Ti-rich region below the Pt cathode (green) and bulk LATP (red).}
  \label{fgr:Oxy_edge}
\end{figure}

To demonstrate that the Ti-rich layer below the Pt cathode is a different phase than the LATP bulk, we performed monochromated STEM-EELS at the two regions (green and red regions) marked in Fig.~\ref{fgr:Oxy_edge} (A). The green region coincide with the Ti-rich layer in the compositional map shown in Fig.~\ref{fgr:comp_mapping} (B). The red region, on the other hand, is considerably deeper inside the LATP cathode and can be treated as representative of the bulk LATP. To compare the local oxygen environments in these two layers, we obtained EELS spectra in the energy region of the Oxygen K-edge, as shown in Fig.~\ref{fgr:Oxy_edge}(B). A small peak is observed at $\approx$ 531 eV from the spectrum obtained from the Ti-rich region, which is nearly absent in the bulk LATP (inset Fig~\ref{fgr:Oxy_edge}). Previous reports have shown that this peak at 531 eV corresponds to amorphous \ce{TiO2} and are absent in bulk LATP. \cite{muto2021stem, gao2014direct}

\subsection{Ab-initio computational investigation of the LATP degradation process}

\begin{figure}
  \includegraphics[width=\textwidth]{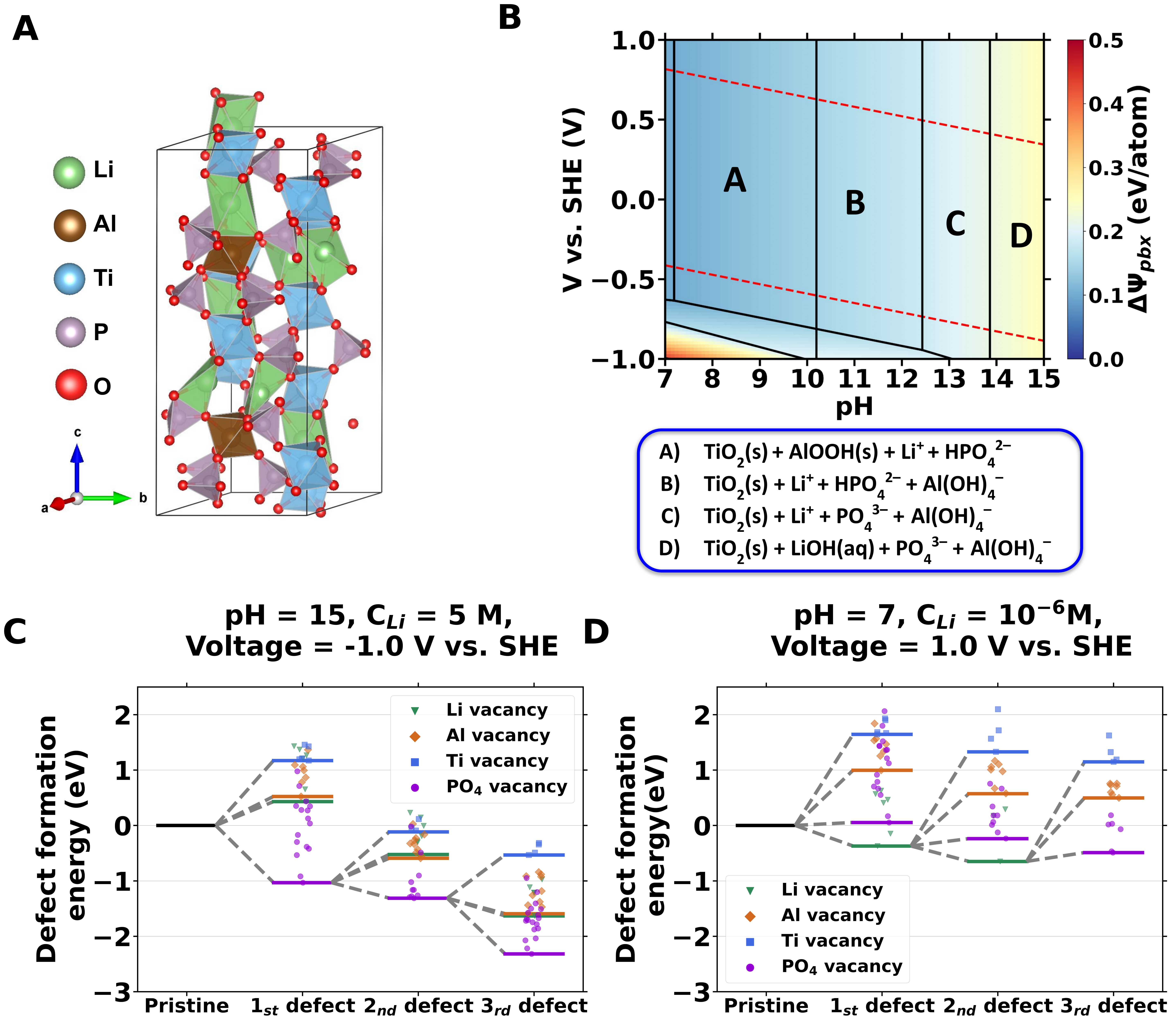}
  \caption{(A) Crystal structure of bulk LATP with composition \ce{Li8Al2Ti10(PO4)18}. (B) \textit{Ab-initio} Pourbaix diagram calculated using R2SCAN-calculated energies for the solids. Concentration of all ions, including Li, are set to $10^{-6}$M. The voltage is referenced to the standard hydrogen electrode (SHE). (C) Surface defect formation energies at high pH, a high Li concentration, and low voltage. (D) Surface defect formation energies at low pH, a low Li concentration, and high voltage. The markers represent the energies of different surface terminations (\ce{H+} or \ce{OH-}) for a given defect type, with the lowest-energy configuration for each defect highlighted by a horizontal line. The results shown in (C) and (D) are calculated with the GGA exchange-correlation functional.}
  \label{fgr:dft}
\end{figure}

The experimental results indicate that LATP undergoes degradation over cycling, leading to the aggregation of degradation products, such as \ce{Li3PO4} and \ce{AlPO4}, on top of the platinum cathode, as well as the enrichment of Ti below the platinum cathode. To understand the energetics of the degradation process, we conduct density functional theory (DFT) calculations to evaluate the stability of LATP under different electrochemical conditions. Figure~\ref{fgr:dft}(A) shows the unit cell of LATP used for the DFT calculations. The chemical formula of this unit cell is \ce{Li8Al2Ti10(PO4)18}, closely resembling the experimental composition.

We first evaluate the thermodynamic bulk stability of LATP under different conditions of pH and applied voltage, from which we construct the Pourbaix diagram as shown in Fig.~\ref{fgr:dft}(B) using the method developed by Persson \textit{et al.}\cite{Persson2012_pbx}. The color scheme in the Pourbaix diagram is used to represent the thermodynamic driving force to decompose LATP into the most stable combinations of solid and/or ionic species in solution. The decomposition products in some major Pourbaix domains are labeled in the blue box. We quantify the driving force of decomposition by the Pourbaix decomposition energy ($\Delta \psi_{pbx}$), which is obtained by taking the difference of the Pourbaix potentials ($\psi_{pbx}$) between LATP and its most stable decomposition products \cite{Singh2017_pbx,Sun2019_pbx}. The Pourbaix potential is the grand potential of a chemical system open to hydrogen, oxygen, and electrons, measuring the thermodynamic stability of a system with a reservoir of liquid water at a given pH and voltage condition. In Fig.~\ref{fgr:dft} (b), we include the range of voltage from -1.0 to 1.0 V vs. the standard hydrogen electrode (SHE), which approximately corresponds to 2.0 to 4.0 V vs. Li/\ce{Li+}. This covers the typical experimental voltage range from 3.0 to 4.5 V vs. Li/\ce{Li+} during the battery operation (see Supplementary Figures S1 and S2). We consider pH values ranging from a neutral ($\text{pH} = 7$) to a highly basic condition ($\text{pH} = 15$). The upper limit of $\text{pH}=15$ is determined by the saturation concentration of LiOH ($C_{Li}\approx$ 5 M) in water at room temperature as LiOH is the reaction product in humidified oxygen \cite{monnin2005thermodynamics,kim2023operando}. Clearly, LATP is not the thermodynamically stable phase, and thus can decompose across all considered pH and voltage conditions, as evidenced by the positive $\Delta \psi_{pbx}$ values. Moreover, LATP becomes more unstable as pH increases, indicated by the more positive $\Delta \psi_{pbx}$ values. 

The Pourbaix diagram also suggests that decomposition products vary under different pH conditions. The phosphate group in LATP always dissolves as ions, either in the form of \ce{HPO4^{2-}} at $\text{pH} < 12$ or \ce{PO4^{3-}} under more strongly basic conditions. Aluminum is present as a solid product (\ce{AlOOH}) at neutral to weakly basic conditions ($\text{pH} < 10$), while it dissolves as ions (\ce{Al(OH)4^{–}}) at higher alkalinity. The lithium easily dissolves as \ce{Li+} ions until the pH increases to $\sim$14, where aqueous LiOH(aq) starts to form. Unlike other elements, the titanium does not dissolve as ions, and forms a \ce{TiO2}(s) solid product across all pH and voltage conditions. Our results indicate that Li, Al, and P have tendency to dissolve as ionic species, whereas Ti is more likely to remain in the solid phase under the conditions relevant to the operation of humidified Li-\ce{O2} battery.

To separate the energetic contributions from each constituting element to the decomposition reaction, we further calculate the energy to remove \ce{Li}, \ce{Al}, \ce{Ti}, or \ce{PO4} from the surface and put them in solution. Compensation of the charge removed is done by  adding \ce{H+} or \ce{OH-} surface terminations (see Methods). The surface vacancies are created one by one up to 3 defects per surface unit cell. More specifically, starting from the pristine (012) LATP surface, which was previously estimated to be the lowest energy surface orientation\cite{Tian2019_latp_surface,Pogosova2020_latp_surface,Lam2024_ltgp_degradation}, we separately create each type of surface vacancy, which is referred to as the 1st defect. The surface with the lowest formation energy from the 1st defect is then used as the initial structure to create a 2nd defect. This process is repeated until a 3rd defect is created.

The surface vacancy formation energies are calculated at different external conditions that LATP electrolyte might experience during the cycling, and are shown in Figs.~\ref{fgr:dft} (C) and (D).  Different data points for each type of surface vacancy correspond to different configurations of surface termination (\ce{H+} or \ce{OH-}), and the lowest energy configuration is highlighted with the horizontal line. The condition used in Fig.~\ref{fgr:dft} (C) is at high pH, a high lithium concentration in solution, and low voltage, which resembles the highly alkaline environment created by the discharge product LiOH. As can be seen, the \ce{PO4} vacancy exhibits the most negative driving force ($\approx -1.0$ eV), a trend that persists up to the formation of the 3rd defect. Similar results are also observed for \ce{LiTi2(PO4)3} and \ce{LiGe2(PO4)3} at high pH condition (pH = 12) in our previous work\cite{Lam2024_ltgp_degradation}, indicating that phosphate group dissolution is a common cause for the degradation of phosphate-based NASICON under highly alkaline conditions. In comparison, the formation of other cation vacancies are thermodynamically unfavorable on the pristine surface, as they have a positive defect formation energy. The same trend persists when the cation vacancies are introduced as the 2nd defect after the first phosphate group is dissolved, as indicated by the uphill energy landscape. However, the energy landscape becomes downhill for Li and Al vacancies after two phosphate groups, corresponding to half of surface phosphate groups per unit surface cell, are dissolved from the surface. This suggests that the pre-existence of surface defects (i.e., \ce{PO4} vacancies) can facilitate the dissolution of other cation species (i.e., \ce{Li} and \ce{Al}). On the other hand, the titanium remains highly resistive to the alkaline corrosion, exhibiting a high thermodynamic cost for dissolution ($\approx$ 0.8 eV) even after half of (two out of four) phosphate groups on the surface are removed.

In Fig.~\ref{fgr:dft} (D), we calculate the energy landscape for the surface dissolution at neutral pH, a low Li concentration, and high voltage, simulating conditions similar to the end of charge, when the discharge product LiOH is fully decomposed and \ce{Li+} is transported back to the anode. Unlike at the high pH condition, the dissolution of phosphate group is no longer thermodynamically favorable, exhibiting a slightly positive energy cost for its defect formation. Our calculations indicate that all lithium on the surface (two lithium per surface unit cell in total) tend to dissolve under this neutral pH condition. No further surface vacancy formations are energetically allowed after the lithium dissolution, as indicated by the uphill energy landscapes for forming all other types of vacancy as the 3rd defect.

\section{Discussion}

\begin{figure}
  \includegraphics[width=\textwidth]{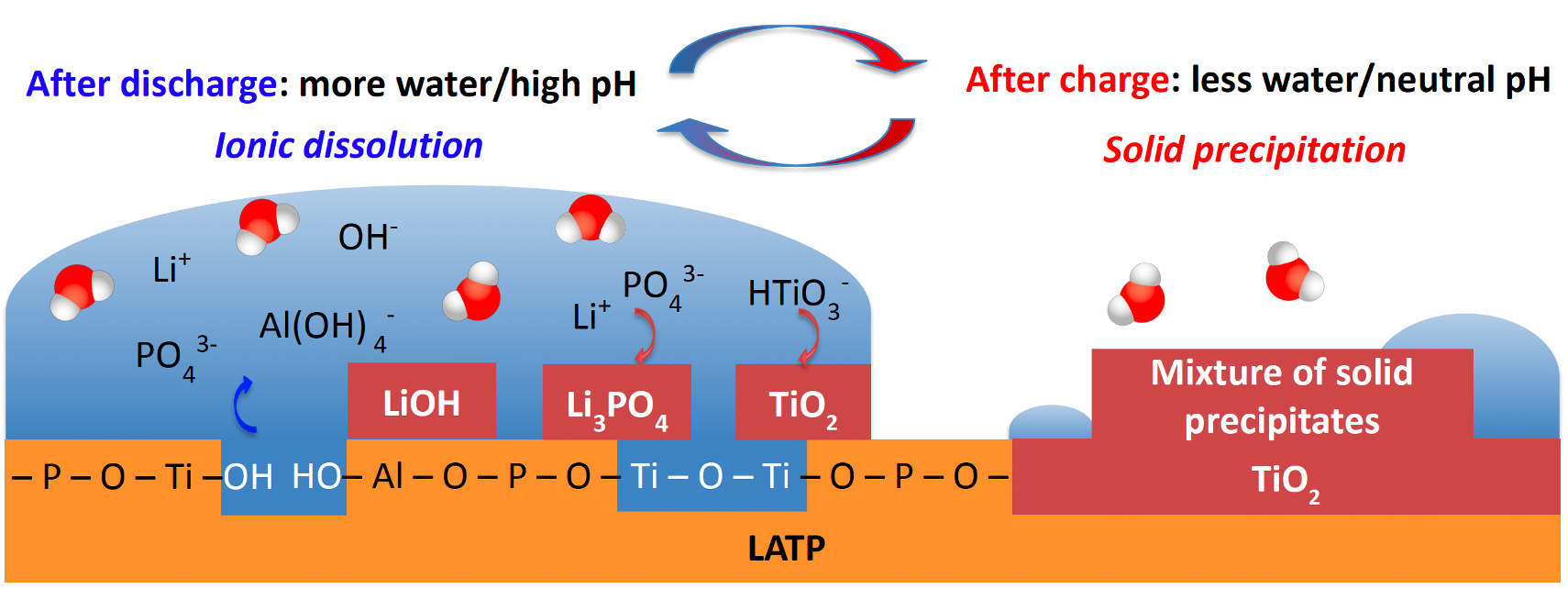}
  \caption{Schematic of LATP degradation mechanism during battery cycling}
  \label{fgr:schematic}
\end{figure}

In this work, we conduct a detailed characterization of LATP degradation after cycling in a humidified \ce{Li-O2} battery. Based on insights from our XPS and EELS characterizations, and DFT calculations of the LATP solid electrolyte at different cycling stages, the degradation mechanism of the LATP solid electrolyte is schematically illustrated in Fig.~\ref{fgr:schematic}.

Previous reports have shown that during cycling, the discharge of the humidified \ce{Li-O2} cell results in the formation of bubble-like discharge products of hydrated \ce{LiOH} \cite{kim2023operando}. Our current experiment also shows that in the initial discharge the primary discharge products are \ce{LiOH} and \ce{Li2O2} (Fig.~\ref{fgr:xps_chara}(A)). Due to the presence of water in the cathode, these discharge products create an extremely alkaline environment (pH $\approx$ 15) on the LATP surface. Our computational results indicate that such a highly alkaline environment leads to the degradation of the LATP electrolyte by first leaching out the \ce{PO4^{3-}} groups, followed by the leaching of \ce{Al^{3+}} ions, from the LATP solid electrolyte surface (Fig.~\ref{fgr:dft}(C)). 

Upon charging, as the discharge product \ce{LiOH} decomposes and the \ce{Li+} ions migrate back to the anode, the pH in the cathode decreases. According to the computational prediction, the surface dissolution is mostly prohibited at neutral pH, which should result in the precipitation of the dissolved ionic species formed during discharging. Furthermore, previous experimental results indicate that the water droplets shrink and begin to disappear \cite{kim2023operando}, forcing the dissolved ions to precipitate as solid products on top of the LATP solid electrolyte. The precipitation of leached \ce{PO4^{3-}} and \ce{Al^{3+}} in subsequent cycles is also confirmed by the SEM surface images (Fig.~\ref{fgr:sem_surf} C,D). Compounds like \ce{AlPO4} and \ce{Li3PO4} are deposited during these solid precipitations, confirmed by XPS measurements (Fig.~\ref{fgr:xps_chara} C, D). Furthermore, the presence of the Al rich corrosion products on top of the LATP electrode is also seen in the cross-sectional compositional mapping using EELS measurements (Fig.~\ref{fgr:comp_mapping}(B)). From these characterization measurements we deduce that the formation of degradation products, such as \ce{AlPO4} and \ce{Li3PO4}, on top of LATP electrolyte and Pt cathode surfaces is a result of battery cycling, where the surface conditions repetitively varies between neutral and high pH, as well as dry and aqueous conditions.

Our DFT calculations show that the Ti defect formation energy is the highest among all defect types in both alkaline and neutral conditions. Therefore, Ti is most likely to remain behind at the LATP surface upon battery cycling, resulting in the formation of a Ti-rich layer right below the Pt cathode layer. This computational result is validated by the composition mapping using EELS (Fig.~\ref{fgr:comp_mapping}) and the change of the Oxygen environment in this layer (Fig.~\ref{fgr:Oxy_edge}) matching well with the \ce{TiO2} standards \cite{muto2021stem, gao2014direct}.

Our observations indicate that within a few cycles a Ti-rich layer forms at the surface of the LATP electrolyte as a result of the leaching of \ce{PO4^{3-}} and \ce{Al^{3+}} ions under alkaline conditions. This observation is consistent with previous reports describing the behavior of LATP powders and pellets dipped in high pH solutions \cite{Lam2024_ltgp_degradation}. The occurrence of side reactions is also reflected in our cycling data, which show that the Coulombic efficiency (CE, defined as the ratio of charge capacity to discharge capacity) of the \ce{Li-O2} battery is only $\approx$ 80\% during the initial cycles, but gradually increases to nearly 100\% after about 10 cycles (supplementary Figure S3). We attribute such improvement in the CE to the reduction of side reactions by the formation of a passivation \ce{TiO2} layer beneath the Pt cathode during the first few cycles. Previous studies on all-solid-state \ce{Li-O2} batteries employing LATP cathodes coated with silicone oil have demonstrated nearly 100\% coulombic efficiency (CE) from the very first cycle \cite{zhu2015high}. The silicone oil film effectively prevents water molecules from reaching the LATP surface, thereby inhibiting the formation of LiOH as discharge product and its dissolution in water. As a result, the high CE can likely be attributed to the absence of a highly alkaline solution environment on the cathode-side of the LATP surface. In fact, it is well-known that \ce{TiO2} has high durability in alkaline environment and its effectiveness as a passivation layer has also been demonstrated \cite{Tio2019_TiO2,Acevedo-pena2013_TiO2,arunachalam2023reliable}.

In this light, the effectiveness of the in-situ formed \ce{TiO2} layer as a protective passivation barrier against the highly alkaline environment generated during cycling in humidified \ce{Li-O2} batteries remains debatable. Indeed, we observe battery cell failure upon extended cycling (\textgreater 49 cycles) under humidified conditions (supplementary Figure S4). A fundamental issue associated with the in-situ formation of the Ti-rich layer is that it inherently involves substantial degradation of the LATP electrolyte surface through the leaching of \ce{PO4^{3-}} and \ce{Al^{3+}} ions. Furthermore, the in-situ formed \ce{TiO2} layer does not seem to guarantee uniform protection of the LATP solid electrolyte against the highly alkaline conditions developed during cycling. This limitation is also evident from the lower than 100 \% CE after 10 cycles (supplementary Figure S3) and the noticeable increase in corrosion products observed on the LATP electrolyte surface after 100 cycles, as shown in Fig. \ref{fgr:sem_surf}(D).

Finally, from our characterization results we propose a promising approach for stable cycling of the solid-state humidified \ce{Li-O2} battery: uniformly coating the LATP with a thin layer (on the nanometer scale) of amorphous \ce{TiO2} to provide conformal protection of the LATP pellet against the extremely corrosive environment. This approach may be able to prevent the degradation of the LATP solid electrolyte and avoid the previously observed clogging of the cathode \cite{kim2022carbon,ma2020mixed}.

\section{Conclusion}

In this work, we combined SEM, STEM-EELS, XPS measurements, and DFT calculations to elucidate the degradation mechanism of the LATP electrolyte in all-solid-state humidified \ce{Li-O2} batteries. Our findings reveal that during discharge, products such as \ce{Li2O2} and \ce{LiOH} create an extremely high alkaline environment, leading to the leaching of \ce{PO4^{3-}} and \ce{Al^{3+}} ions, leaving behind a thin Ti-rich layer (approximately 20 nm) on the surface of the LATP electrolyte. Upon charging, the dissolved ions precipitate on the LATP surface as \ce{Li3PO4} and \ce{AlPO4}. The formation of a Ti-rich layer during cycling can potentially mitigate parasitic reactions in the high alkaline environment, thereby extending the cycle life of LATP-based \ce{Li-O2} batteries. However, the Ti-rich layer formed during battery cycling is insufficient to fully prevent cell failure over prolonged cycling. These insights pave the way for engineering alkaline-stable solid-state \ce{Li-O2} batteries through approaches such as conformal coatings of thin layers of amorphous \ce{TiO2}.

\section{Method}

\subsection{Humidified \ce{Li-O2} cell cycling}

The humidified \ce{Li-O2} cathode was fabricated by sputtering a 100 $\mu$m Pt pattern onto a solid LATP electrolyte, which was obtained from Ohara Corporation. The cells were assembled as hybrid cells, with pure Li metal used as the anode. An interlayer consisting of Celgard soaked in a propylene carbonate electrolyte containing 1 M LiTFSI was employed. The Li anode, interlayer, and LATP solid electrolyte were stacked in series and sealed in a pouch, with the Pt cathode exposed to humidified \ce{O2} through an opening. A Pt mesh served as the current collector. The cells were cycled between 3 and 4.5 V using a constant current step of 10 
$\mu$A, with a cutoff voltage of 4.5 V during charging and a cutoff capacity of 0.01 mAh during discharging. Three cells were disassembled after 1, 30, and 100 cycles, and the LATP solid electrolyte along with the Pt cathode was extracted for advanced characterization. For the calculation of the specific capacity, the obtained capacity was normalized with the weight of the sputtered Pt, which was calculated to be $1.078 \times 10^{-2}$ mg. In Supplementary Figures S3 and S4 we report the coulombic efficiency after 30 and 50 cell cycles respectively, calculated as the charge capacity (oxygen evolution) normalized by the discharge capacity (oxygen reduction).

\subsection{X-ray photoelectron spectroscopy (XPS)}

The XPS  data was obtained in a Thermo-Fisher k-Alpha Plus X-ray photoelectron spectrometer at the Molecular Foundry - LBNL using Al K $\alpha$ (1486.6 eV) radiation. The sample pellets were fixed in a sample holder and inserted into the UHV chamber. The photoelectrons were collected using a hemispherical analyzer (HSA) operating in the CAE mode. During the measurements, an electron flood source was used for neutralization of the samples. The measurements were performed in the Survey, Li-1s, P-2p, Al-2p, Ti-2p, O-1s, C-1s, Ge-3d and Si-2p  regions with an energy step size of 1 eV and 0.1 eV  and pass energy of 200 eV and 50 eV for the Survey and high-resolution measurements, respectively.
KOLXPD software (v. 1.8.0) was used for the XPS analysis. For the fitting of the high-resolution spectra, a Shirley-type background was used. The Gaussian width of the peaks was fixed at the same value for all the peaks. The Lorentzian width was fixed for the same components. C 1s at 284.6 eV was used for the calibration of the XPS spectra (Supplementary Fig.S7). The Li components were confirmed by the difference of binding energy between the Li 1s and O 1s peaks, as shown in Ref.\cite{wood2018xps}. 

\subsection{Transmission electron microscopy sample preparation}

To prepare the samples for cross-sectional TEM measurements, a thin lamella from the LATP solid electrolyte with the Pt cathode was meticulously made using the dual-beam focused ion beam method, using a Helios G4 UX instrument. To prevent surface damage a 200 nm thick layer of Pt was deposited on the top of the sample using electron beam followed by a 1.5 $\mu$m thick Pt layer deposited using the Ga-ion beam. The cross-sectional lamella was thinned down to $\approx$ 1 $\mu$m using a Ga-ion beam of 30 kV accelerating voltage and 0.75 nA current. To prevent ion beam implantation subsequent thinning to $\approx$ 500 nm was performed using a 5 kV accelerating voltage and 0.15 nA current. This was followed by a thinning to $\approx$ 100 nm using a 2 kV ion beam at 44 pA current. To prevent carbon contamination of the sample a plasma cleaning of 30 seconds was performed right before the TEM measurements.

\subsection{Electron energy loss spectroscopy measurement}

The electron energy loss measurements for the compositional mapping were performed using the TEAM 1 microscope (double-aberration-corrected Thermo Fisher Scientific Titan 80-300) at the National Center for Electron Microscopy. The EELS measurement was collected using a Gatan Continuum spectrometer at 300 kV, using a convergence angle of approximately 40 mrad and a collection angle of 110 mrad. The width of the zero loss peak was measured to be 1.3 eV and a 0.18 eV per channel dispersion was used to collect the spectra. The compositional maps were drawn using the Gatan digital micrograph software v3.6.  

Similar settings were used to obtain the changes in the oxygen bonding environment with the addition of a monochromator. The addition of the  monochromator reduced the FWHM of the zero loss peak to 0.46 eV.  

\subsection{Computational modeling}

All DFT calculations were performed using
the Vienna Ab-Initio Software Package (VASP)\cite{Kresse1996_vasp} within the projector augmented wave (PAW) formalism\cite{Kresse1996_paw}.Pourbaix diagrams were constructed from ab-initio energies obtained with the R2SCAN functional\cite{Bartok2019_r2scan} using the formalism developed by Persson et al.\cite{Persson2012_pbx}. Plane-wave basis cutoff energies were set at 680 eV. For the surface defect energy calculations, we used the Perdew-Burke-Ernzerhof (PBE) generalized gradient approximation (GGA) functional\cite{Perdew1996_pbe} for higher computational efficiency. Plane-wave basis cutoff energies were set at 520 eV. The formation energies of oxides were corrected with the anion correction scheme implemented in pymatgen\cite{Ong2013_pymatgen} to account for the GGA error that overbinds the \ce{O2} molecule. Electronic and ionic optimization convergence criteria for all calculations, including both r2SCAN and PBE calculations, are set to 10E-5 eV and 0.02 eV/Å, respectively. Other INCAR parameters and k-point grid were generated by MPScanRelaxSet and MPRelaxSet implemented in pymatgen \cite{Ong2013_pymatgen}.

The lattice constants of slabs were first determined by fully relaxing the bulk structure of LATP (\ce{Li8Al2Ti10(PO4)18}) at the PBE level. The surface unit cell is created by transforming the bulk from the conventional hexagonal cell with R-3c symmetry to the orthorhombic cell, where the planes normal to the a, b, and c axes of the orthorhombic cell are the (0–14), (2-10), and (012) planes of the hexagonal cell, respectively. Stoichiometric and symmetric surface slabs were then created with 18 Å slab thickness and a 15 Å vacuum layer. The surface energy of the pristine surface without any surface defects was obtained by fully relaxing atomic positions while fixing the in-plane lattice constants. Defected surfaces were created by removing or adding atoms symmetrically from both the top and bottom surfaces of the pristine slabs. The atomic positions of outer layers (5–6 Å thick layer on each side of a slab) were fully relaxed for defective surface slabs while keeping the rest of the inner-layer atoms fixed at the same positions as in the pristine surfaces.

Surface defect formation energies were calculated as follows:
\begin{equation}
    \Delta E_{\text{defect}} = (E_{\text{d-surface}} - \Delta N_H \times E_H - \Delta N_O \times E_O - \Delta N_{\text{ion}} \times E_{\text{ion}} - E_{\text{p-surface}})/2
\end{equation}
where $\Delta E_{\text{defect}}$ is surface defect formation energy, $\Delta E_{\text{p-surface}}$ is the DFT energy of a pristine surface, and $\Delta E_{\text{d-surface}}$ is DFT energy of a defected surface. $\Delta N_H$, $\Delta N_O$,and $\Delta N_{\text{ion}}$, are the number of hydrogen, oxygen, and ions, respectively, added or removed on both sides of the slab to form the surface defects. A positive value indicates the addition of extra atoms, while a negative value indicates removal. The energies for hydrogen ($E_H$) and oxygen ($E_O$) are corrected following the methodology of Persson et al. \cite{Persson2012_pbx}, where $E_H$ and $E_O$ are adjusted to reproduce the experimental formation energy of liquid water: $\Delta E_{f,H_2O}=E_{H_2O}-2E_H-E_O=–2.4583$ eV). The pH and voltage dependence is incorporated into the hydrogen energy as $E_H = E_H^{\text{ref}} - k_B T \ln10\cdot \text{pH} - eV$, where $E_H^{\text{ref}}$ is the reference hydrogen energy under standard condition (pH = 0, $V$ = 0 V vs. SHE). In practice, $E_{H_2O}$ and $E_{O_2}$ are calculated as the 0 K DFT energies corrected by adding the experimental room-temperature entropies of liquid water and oxygen gas, respectively. The value of $E_H^{\text{ref}}$ is then fitted to ensure the experimental water formation energy is obtained. The energy of removed surface species ($E_{\text{ion}}$) is determined based on experimental formation energy of the most stable ionic species of the corresponding element in solution at given pH and voltage conditions \cite{Sun2019_pbx}. The value of $E_{\text{ion}}$ is corrected by the difference between the experimental and the DFT-calculated energies of pre-selected reference solid phase \cite{Persson2012_pbx}. The dependence of $E_{\text{ion}}$ on concentration and applied voltage is also considered by adding an additional term of $+k_B T \ln [C_{ion}]-QV$, where $C_{ion}$ and $Q$ are the concentration and the charge of the ionic species, respectively. All the energy correction schemes mentioned above are implemented in pymatgen \cite{Ong2013_pymatgen}. The factor of 1/2 is due to the symmetrically formed defects on both sides of the slab. All surface vacancies are charge-compensated by terminating the dangling surface oxygen or the undercoordinated cation by \ce{H+} or \ce{OH-}, respectively, which was shown in recent work \cite{Lam2024_ltgp_degradation} to result in lower energy than that without the termination.


\begin{acknowledgement}
\textbf{Funding:} This work was supported by the Samsung Advanced Institute of Technology (SAIT), Samsung Electronics, Co., Ltd.. The computational analysis was performed using computational resources sponsored by the Department of Energy’s Office of Energy Efficiency and Renewable Energy located at the National Renewable Energy Laboratory (NREL). Computational resources were also provided by the Advanced Cyberinfrastructure Coordination Ecosystem: Services \& Support (ACCESS) program, which is supported by National Science Foundation grants \#2138259, \#2138286, \#2138307, \#2137603, and \#2138296. Work at the Molecular Foundry was supported by the Office of Science, Office of Basic Energy Sciences, of the U.S. Department of Energy under Contract No. DE-AC02-05CH11231. \textbf{Author contributions:} Conceptualization: TPM and ZL. Methodology: TPM and ZL. Electron Microscopy Characterization: TPM, MS, and KB. Computation: ZL. XPS characterization: MJ and LPM. Investigation: TPM, ZL, MS, MJ, LPM, JOP, HK. Visualization: TPM and ZL. Supervision: GC and MS. Writing-original draft: TPM and ZL. Writing-review and editing: All authors. \textbf{Competing interests: } The authors declare that they have no competing interests. \textbf{Data and materials availability:} All  data needed to evaluate the conclusions in the paper are present in the paper and/or the  Supplementary Materials. 
\end{acknowledgement}




\bibliographystyle{achemso}
\bibliography{achemso-demo}

\providecommand{\latin}[1]{#1}
\makeatletter
\providecommand{\doi}
  {\begingroup\let\do\@makeother\dospecials
  \catcode`\{=1 \catcode`\}=2 \doi@aux}
\providecommand{\doi@aux}[1]{\endgroup\texttt{#1}}
\makeatother
\providecommand*\mcitethebibliography{\thebibliography}
\csname @ifundefined\endcsname{endmcitethebibliography}
  {\let\endmcitethebibliography\endthebibliography}{}
\begin{mcitethebibliography}{79}
\providecommand*\natexlab[1]{#1}
\providecommand*\mciteSetBstSublistMode[1]{}
\providecommand*\mciteSetBstMaxWidthForm[2]{}
\providecommand*\mciteBstWouldAddEndPuncttrue
  {\def\EndOfBibitem{\unskip.}}
\providecommand*\mciteBstWouldAddEndPunctfalse
  {\let\EndOfBibitem\relax}
\providecommand*\mciteSetBstMidEndSepPunct[3]{}
\providecommand*\mciteSetBstSublistLabelBeginEnd[3]{}
\providecommand*\EndOfBibitem{}
\mciteSetBstSublistMode{f}
\mciteSetBstMaxWidthForm{subitem}{(\alph{mcitesubitemcount})}
\mciteSetBstSublistLabelBeginEnd
  {\mcitemaxwidthsubitemform\space}
  {\relax}
  {\relax}

\bibitem[Lee \latin{et~al.}(2011)Lee, Tai~Kim, Cao, Choi, Liu, Lee, and
  Cho]{lee2011metal}
Lee,~J.-S.; Tai~Kim,~S.; Cao,~R.; Choi,~N.-S.; Liu,~M.; Lee,~K.~T.; Cho,~J.
  Metal--air batteries with high energy density: Li--air versus Zn--air.
  \emph{Advanced Energy Materials} \textbf{2011}, \emph{1}, 34--50\relax
\mciteBstWouldAddEndPuncttrue
\mciteSetBstMidEndSepPunct{\mcitedefaultmidpunct}
{\mcitedefaultendpunct}{\mcitedefaultseppunct}\relax
\EndOfBibitem
\bibitem[Capasso and Veneri(2014)Capasso, and Veneri]{capasso2014experimental}
Capasso,~C.; Veneri,~O. Experimental analysis on the performance of lithium
  based batteries for road full electric and hybrid vehicles. \emph{Applied
  Energy} \textbf{2014}, \emph{136}, 921--930\relax
\mciteBstWouldAddEndPuncttrue
\mciteSetBstMidEndSepPunct{\mcitedefaultmidpunct}
{\mcitedefaultendpunct}{\mcitedefaultseppunct}\relax
\EndOfBibitem
\bibitem[Au \latin{et~al.}(2022)Au, Crespo-Ribadeneyra, and
  Titirici]{au2022beyond}
Au,~H.; Crespo-Ribadeneyra,~M.; Titirici,~M.-M. Beyond Li-ion batteries:
  performance, materials diversification, and sustainability. \emph{One Earth}
  \textbf{2022}, \emph{5}, 207--211\relax
\mciteBstWouldAddEndPuncttrue
\mciteSetBstMidEndSepPunct{\mcitedefaultmidpunct}
{\mcitedefaultendpunct}{\mcitedefaultseppunct}\relax
\EndOfBibitem
\bibitem[Zheng \latin{et~al.}(2008)Zheng, Liang, Hendrickson, and
  Plichta]{zheng2008theoretical}
Zheng,~J.; Liang,~R.; Hendrickson,~M.~a.; Plichta,~E. Theoretical energy
  density of Li--air batteries. \emph{Journal of the Electrochemical Society}
  \textbf{2008}, \emph{155}, A432\relax
\mciteBstWouldAddEndPuncttrue
\mciteSetBstMidEndSepPunct{\mcitedefaultmidpunct}
{\mcitedefaultendpunct}{\mcitedefaultseppunct}\relax
\EndOfBibitem
\bibitem[Tan \latin{et~al.}(2013)Tan, Wei, Shyy, and Zhao]{tan2013prediction}
Tan,~P.; Wei,~Z.; Shyy,~W.; Zhao,~T. Prediction of the theoretical capacity of
  non-aqueous lithium-air batteries. \emph{Applied energy} \textbf{2013},
  \emph{109}, 275--282\relax
\mciteBstWouldAddEndPuncttrue
\mciteSetBstMidEndSepPunct{\mcitedefaultmidpunct}
{\mcitedefaultendpunct}{\mcitedefaultseppunct}\relax
\EndOfBibitem
\bibitem[Ma \latin{et~al.}(2015)Ma, Yuan, Li, Ma, Wilkinson, Zhang, and
  Zhang]{Ma2015_review}
Ma,~Z.; Yuan,~X.; Li,~L.; Ma,~Z.-f.; Wilkinson,~D.~P.; Zhang,~L.; Zhang,~J. {A
  review of cathode materials and structures for rechargeable lithium–air
  batteries}. \emph{Energy \& Environmental Science} \textbf{2015}, \emph{8},
  2144\relax
\mciteBstWouldAddEndPuncttrue
\mciteSetBstMidEndSepPunct{\mcitedefaultmidpunct}
{\mcitedefaultendpunct}{\mcitedefaultseppunct}\relax
\EndOfBibitem
\bibitem[Luntz and McCloskey(2014)Luntz, and McCloskey]{luntz2014nonaqueous}
Luntz,~A.~C.; McCloskey,~B.~D. Nonaqueous Li--air batteries: a status report.
  \emph{Chemical reviews} \textbf{2014}, \emph{114}, 11721--11750\relax
\mciteBstWouldAddEndPuncttrue
\mciteSetBstMidEndSepPunct{\mcitedefaultmidpunct}
{\mcitedefaultendpunct}{\mcitedefaultseppunct}\relax
\EndOfBibitem
\bibitem[Tian \latin{et~al.}(2021)Tian, Zeng, Rutt, Shi, Kim, Wang, Koettgen,
  Sun, Ouyang, Chen, Lun, Rong, Persson, and Ceder]{Tian2021_review}
Tian,~Y.; Zeng,~G.; Rutt,~A.; Shi,~T.; Kim,~H.; Wang,~J.; Koettgen,~J.;
  Sun,~Y.; Ouyang,~B.; Chen,~T.; Lun,~Z.; Rong,~Z.; Persson,~K.; Ceder,~G.
  {Promises and Challenges of Next-Generation "beyond Li-ion" Batteries for
  Electric Vehicles and Grid Decarbonization}. \emph{Chemical Reviews}
  \textbf{2021}, \emph{121}, 1623--1669\relax
\mciteBstWouldAddEndPuncttrue
\mciteSetBstMidEndSepPunct{\mcitedefaultmidpunct}
{\mcitedefaultendpunct}{\mcitedefaultseppunct}\relax
\EndOfBibitem
\bibitem[Lu \latin{et~al.}(2010)Lu, Gasteiger, Parent, Chiloyan, and
  Shao-Horn]{lu2010influence}
Lu,~Y.-C.; Gasteiger,~H.~A.; Parent,~M.~C.; Chiloyan,~V.; Shao-Horn,~Y. The
  influence of catalysts on discharge and charge voltages of rechargeable
  Li--oxygen batteries. \emph{Electrochemical and Solid-State Letters}
  \textbf{2010}, \emph{13}, A69\relax
\mciteBstWouldAddEndPuncttrue
\mciteSetBstMidEndSepPunct{\mcitedefaultmidpunct}
{\mcitedefaultendpunct}{\mcitedefaultseppunct}\relax
\EndOfBibitem
\bibitem[McCloskey \latin{et~al.}(2011)McCloskey, Bethune, Shelby, Girishkumar,
  and Luntz]{McCloskey2011_solvent}
McCloskey,~B.~D.; Bethune,~D.~S.; Shelby,~R.~M.; Girishkumar,~G.; Luntz,~A.~C.
  {Solvents critical role in nonaqueous Lithium-Oxygen battery
  electrochemistry}. \emph{Journal of Physical Chemistry Letters}
  \textbf{2011}, \emph{2}, 1161--1166\relax
\mciteBstWouldAddEndPuncttrue
\mciteSetBstMidEndSepPunct{\mcitedefaultmidpunct}
{\mcitedefaultendpunct}{\mcitedefaultseppunct}\relax
\EndOfBibitem
\bibitem[Freunberger \latin{et~al.}(2011)Freunberger, Chen, Peng, Griffin,
  Hardwick, Bard{\'e}, Nov{\'a}k, and Bruce]{freunberger2011reactions}
Freunberger,~S.~A.; Chen,~Y.; Peng,~Z.; Griffin,~J.~M.; Hardwick,~L.~J.;
  Bard{\'e},~F.; Nov{\'a}k,~P.; Bruce,~P.~G. Reactions in the rechargeable
  lithium--O2 battery with alkyl carbonate electrolytes. \emph{Journal of the
  American Chemical Society} \textbf{2011}, \emph{133}, 8040--8047\relax
\mciteBstWouldAddEndPuncttrue
\mciteSetBstMidEndSepPunct{\mcitedefaultmidpunct}
{\mcitedefaultendpunct}{\mcitedefaultseppunct}\relax
\EndOfBibitem
\bibitem[McCloskey \latin{et~al.}(2012)McCloskey, Speidel, Scheffler, Miller,
  Viswanathan, Hummelsh{\o}j, N{\o}rskov, and Luntz]{mccloskey2012twin}
McCloskey,~B.~D.; Speidel,~A.; Scheffler,~R.; Miller,~D.; Viswanathan,~V.;
  Hummelsh{\o}j,~J.; N{\o}rskov,~J.; Luntz,~A. Twin problems of interfacial
  carbonate formation in nonaqueous Li--O2 batteries. \emph{The journal of
  physical chemistry letters} \textbf{2012}, \emph{3}, 997--1001\relax
\mciteBstWouldAddEndPuncttrue
\mciteSetBstMidEndSepPunct{\mcitedefaultmidpunct}
{\mcitedefaultendpunct}{\mcitedefaultseppunct}\relax
\EndOfBibitem
\bibitem[Ottakam~Thotiyl \latin{et~al.}(2013)Ottakam~Thotiyl, Freunberger,
  Peng, and Bruce]{ottakam2013carbon}
Ottakam~Thotiyl,~M.~M.; Freunberger,~S.~A.; Peng,~Z.; Bruce,~P.~G. The carbon
  electrode in nonaqueous Li--O2 cells. \emph{Journal of the American Chemical
  Society} \textbf{2013}, \emph{135}, 494--500\relax
\mciteBstWouldAddEndPuncttrue
\mciteSetBstMidEndSepPunct{\mcitedefaultmidpunct}
{\mcitedefaultendpunct}{\mcitedefaultseppunct}\relax
\EndOfBibitem
\bibitem[Hong(1978)]{hong1978crystal}
Hong,~H.-P. Crystal structure and ionic conductivity of Li14Zn (GeO4) 4 and
  other new Li+ superionic conductors. \emph{Materials Research Bulletin}
  \textbf{1978}, \emph{13}, 117--124\relax
\mciteBstWouldAddEndPuncttrue
\mciteSetBstMidEndSepPunct{\mcitedefaultmidpunct}
{\mcitedefaultendpunct}{\mcitedefaultseppunct}\relax
\EndOfBibitem
\bibitem[Adachi \latin{et~al.}(1996)Adachi, Imanaka, and Aono]{adachi1996fast}
Adachi,~G.-y.; Imanaka,~N.; Aono,~H. Fast Li+ conducting ceramic electrolytes.
  \emph{Advanced Materials} \textbf{1996}, \emph{8}, 127--135\relax
\mciteBstWouldAddEndPuncttrue
\mciteSetBstMidEndSepPunct{\mcitedefaultmidpunct}
{\mcitedefaultendpunct}{\mcitedefaultseppunct}\relax
\EndOfBibitem
\bibitem[MacFarlane \latin{et~al.}(1999)MacFarlane, Huang, and
  Forsyth]{macfarlane1999lithium}
MacFarlane,~D.~R.; Huang,~J.; Forsyth,~M. Lithium-doped plastic crystal
  electrolytes exhibiting fast ion conduction for secondary batteries.
  \emph{Nature} \textbf{1999}, \emph{402}, 792--794\relax
\mciteBstWouldAddEndPuncttrue
\mciteSetBstMidEndSepPunct{\mcitedefaultmidpunct}
{\mcitedefaultendpunct}{\mcitedefaultseppunct}\relax
\EndOfBibitem
\bibitem[Knauth(2009)]{knauth2009inorganic}
Knauth,~P. Inorganic solid Li ion conductors: An overview. \emph{Solid State
  Ionics} \textbf{2009}, \emph{180}, 911--916\relax
\mciteBstWouldAddEndPuncttrue
\mciteSetBstMidEndSepPunct{\mcitedefaultmidpunct}
{\mcitedefaultendpunct}{\mcitedefaultseppunct}\relax
\EndOfBibitem
\bibitem[Chi \latin{et~al.}(2021)Chi, Li, Di, Bai, Song, Wang, Li, Liang, Xu,
  and Yu]{chi2021highly}
Chi,~X.; Li,~M.; Di,~J.; Bai,~P.; Song,~L.; Wang,~X.; Li,~F.; Liang,~S.;
  Xu,~J.; Yu,~J. A highly stable and flexible zeolite electrolyte solid-state
  Li--air battery. \emph{Nature} \textbf{2021}, \emph{592}, 551--557\relax
\mciteBstWouldAddEndPuncttrue
\mciteSetBstMidEndSepPunct{\mcitedefaultmidpunct}
{\mcitedefaultendpunct}{\mcitedefaultseppunct}\relax
\EndOfBibitem
\bibitem[Le \latin{et~al.}(2019)Le, Ngo, Didwal, Fisher, Park, Kim, and
  Park]{le2019highly}
Le,~H.~T.; Ngo,~D.~T.; Didwal,~P.~N.; Fisher,~J.~G.; Park,~C.-N.; Kim,~I.-D.;
  Park,~C.-J. Highly efficient and stable solid-state Li--O 2 batteries using a
  perovskite solid electrolyte. \emph{Journal of Materials Chemistry A}
  \textbf{2019}, \emph{7}, 3150--3160\relax
\mciteBstWouldAddEndPuncttrue
\mciteSetBstMidEndSepPunct{\mcitedefaultmidpunct}
{\mcitedefaultendpunct}{\mcitedefaultseppunct}\relax
\EndOfBibitem
\bibitem[Kim \latin{et~al.}(2022)Kim, Lee, Kwon, Bak, Jaye, Fischer, Yoon,
  Park, Seo, Ma, \latin{et~al.} others]{kim2022carbon}
Kim,~M.; Lee,~H.; Kwon,~H.~J.; Bak,~S.-M.; Jaye,~C.; Fischer,~D.~A.; Yoon,~G.;
  Park,~J.~O.; Seo,~D.-H.; Ma,~S.~B.; others Carbon-free high-performance
  cathode for solid-state Li-O2 battery. \emph{Science Advances} \textbf{2022},
  \emph{8}, eabm8584\relax
\mciteBstWouldAddEndPuncttrue
\mciteSetBstMidEndSepPunct{\mcitedefaultmidpunct}
{\mcitedefaultendpunct}{\mcitedefaultseppunct}\relax
\EndOfBibitem
\bibitem[Radin and Siegel(2013)Radin, and Siegel]{Radin2013_charge}
Radin,~M.~D.; Siegel,~D.~J. {Charge transport in lithium peroxide: Relevance
  for rechargeable metal-air batteries}. \emph{Energy and Environmental
  Science} \textbf{2013}, \emph{6}, 2370--2379\relax
\mciteBstWouldAddEndPuncttrue
\mciteSetBstMidEndSepPunct{\mcitedefaultmidpunct}
{\mcitedefaultendpunct}{\mcitedefaultseppunct}\relax
\EndOfBibitem
\bibitem[Luntz \latin{et~al.}(2013)Luntz, Viswanathan, Voss, Varley,
  N{\o}rskov, Scheffler, and Speidel]{luntz2013tunneling}
Luntz,~A.~C.; Viswanathan,~V.; Voss,~J.; Varley,~J.; N{\o}rskov,~J.;
  Scheffler,~R.; Speidel,~A. Tunneling and polaron charge transport through
  Li2O2 in Li--O2 batteries. \emph{The Journal of Physical Chemistry Letters}
  \textbf{2013}, \emph{4}, 3494--3499\relax
\mciteBstWouldAddEndPuncttrue
\mciteSetBstMidEndSepPunct{\mcitedefaultmidpunct}
{\mcitedefaultendpunct}{\mcitedefaultseppunct}\relax
\EndOfBibitem
\bibitem[Radin \latin{et~al.}(2012)Radin, Tian, and
  Siegel]{radin2012electronic}
Radin,~M.~D.; Tian,~F.; Siegel,~D.~J. Electronic structure of Li 2 O 2
  $\{$0001$\}$ surfaces. \emph{Journal of Materials Science} \textbf{2012},
  \emph{47}, 7564--7570\relax
\mciteBstWouldAddEndPuncttrue
\mciteSetBstMidEndSepPunct{\mcitedefaultmidpunct}
{\mcitedefaultendpunct}{\mcitedefaultseppunct}\relax
\EndOfBibitem
\bibitem[Viswanathan \latin{et~al.}(2011)Viswanathan, Thygesen, Hummelsh{\o}j,
  N{\o}rskov, Girishkumar, McCloskey, and Luntz]{viswanathan2011electrical}
Viswanathan,~V.; Thygesen,~K.~S.; Hummelsh{\o}j,~J.; N{\o}rskov,~J.~K.;
  Girishkumar,~G.; McCloskey,~B.; Luntz,~A. Electrical conductivity in Li2O2
  and its role in determining capacity limitations in non-aqueous Li-O2
  batteries. \emph{The Journal of chemical physics} \textbf{2011},
  \emph{135}\relax
\mciteBstWouldAddEndPuncttrue
\mciteSetBstMidEndSepPunct{\mcitedefaultmidpunct}
{\mcitedefaultendpunct}{\mcitedefaultseppunct}\relax
\EndOfBibitem
\bibitem[Gerbig \latin{et~al.}(2013)Gerbig, Merkle, and
  Maier]{gerbig2013electron}
Gerbig,~O.; Merkle,~R.; Maier,~J. Electron and ion transport in Li2O2.
  \emph{Advanced materials} \textbf{2013}, \emph{25}, 3129--3133\relax
\mciteBstWouldAddEndPuncttrue
\mciteSetBstMidEndSepPunct{\mcitedefaultmidpunct}
{\mcitedefaultendpunct}{\mcitedefaultseppunct}\relax
\EndOfBibitem
\bibitem[Radin \latin{et~al.}(2015)Radin, Monroe, and Siegel]{radin2015dopants}
Radin,~M.~D.; Monroe,~C.~W.; Siegel,~D.~J. How dopants can enhance charge
  transport in Li2O2. \emph{Chemistry of Materials} \textbf{2015}, \emph{27},
  839--847\relax
\mciteBstWouldAddEndPuncttrue
\mciteSetBstMidEndSepPunct{\mcitedefaultmidpunct}
{\mcitedefaultendpunct}{\mcitedefaultseppunct}\relax
\EndOfBibitem
\bibitem[Zhang \latin{et~al.}(2022)Zhang, Xiao, Yu, Zhao, and
  Tan]{zhang2022reacquainting}
Zhang,~Z.; Xiao,~X.; Yu,~W.; Zhao,~Z.; Tan,~P. Reacquainting the sudden-death
  and reaction routes of Li--O2 batteries by ex situ observation of Li2O2
  distribution inside a highly ordered air electrode. \emph{Nano Letters}
  \textbf{2022}, \emph{22}, 7527--7534\relax
\mciteBstWouldAddEndPuncttrue
\mciteSetBstMidEndSepPunct{\mcitedefaultmidpunct}
{\mcitedefaultendpunct}{\mcitedefaultseppunct}\relax
\EndOfBibitem
\bibitem[Liu \latin{et~al.}(2015)Liu, Leskes, Yu, Moore, Zhou, Bayley, Kim, and
  Grey]{liu2015cycling}
Liu,~T.; Leskes,~M.; Yu,~W.; Moore,~A.~J.; Zhou,~L.; Bayley,~P.~M.; Kim,~G.;
  Grey,~C.~P. Cycling Li-O2 batteries via LiOH formation and decomposition.
  \emph{Science} \textbf{2015}, \emph{350}, 530--533\relax
\mciteBstWouldAddEndPuncttrue
\mciteSetBstMidEndSepPunct{\mcitedefaultmidpunct}
{\mcitedefaultendpunct}{\mcitedefaultseppunct}\relax
\EndOfBibitem
\bibitem[Ma \latin{et~al.}(2020)Ma, Kwon, Kim, Bak, Lee, Ehrlich, Cho, Im, and
  Seo]{ma2020mixed}
Ma,~S.~B.; Kwon,~H.~J.; Kim,~M.; Bak,~S.-M.; Lee,~H.; Ehrlich,~S.~N.;
  Cho,~J.-J.; Im,~D.; Seo,~D.-H. Mixed ionic--electronic conductor of
  perovskite LixLayMO3- $\delta$ toward carbon-free cathode for reversible
  lithium--air batteries. \emph{Advanced Energy Materials} \textbf{2020},
  \emph{10}, 2001767\relax
\mciteBstWouldAddEndPuncttrue
\mciteSetBstMidEndSepPunct{\mcitedefaultmidpunct}
{\mcitedefaultendpunct}{\mcitedefaultseppunct}\relax
\EndOfBibitem
\bibitem[Lu \latin{et~al.}(2014)Lu, Li, Park, Sun, Wu, and
  Amine]{lu2014aprotic}
Lu,~J.; Li,~L.; Park,~J.-B.; Sun,~Y.-K.; Wu,~F.; Amine,~K. Aprotic and aqueous
  Li--O2 batteries. \emph{Chemical reviews} \textbf{2014}, \emph{114},
  5611--5640\relax
\mciteBstWouldAddEndPuncttrue
\mciteSetBstMidEndSepPunct{\mcitedefaultmidpunct}
{\mcitedefaultendpunct}{\mcitedefaultseppunct}\relax
\EndOfBibitem
\bibitem[Gao \latin{et~al.}(2023)Gao, Temprano, Lei, Tang, Li, Grey, and
  Liu]{gao2023recent}
Gao,~Z.; Temprano,~I.; Lei,~J.; Tang,~L.; Li,~J.; Grey,~C.~P.; Liu,~T. Recent
  Progress in Developing a LiOH-Based Reversible Nonaqueous Lithium--Air
  Battery. \emph{Advanced Materials} \textbf{2023}, \emph{35}, 2201384\relax
\mciteBstWouldAddEndPuncttrue
\mciteSetBstMidEndSepPunct{\mcitedefaultmidpunct}
{\mcitedefaultendpunct}{\mcitedefaultseppunct}\relax
\EndOfBibitem
\bibitem[Dai \latin{et~al.}(2019)Dai, Li, Liu, Amine, and
  Lu]{dai2019fundamental}
Dai,~A.; Li,~Q.; Liu,~T.; Amine,~K.; Lu,~J. Fundamental Understanding of
  Water-Induced Mechanisms in Li--O2 Batteries: Recent Developments and
  Perspectives. \emph{Advanced Materials} \textbf{2019}, \emph{31},
  1805602\relax
\mciteBstWouldAddEndPuncttrue
\mciteSetBstMidEndSepPunct{\mcitedefaultmidpunct}
{\mcitedefaultendpunct}{\mcitedefaultseppunct}\relax
\EndOfBibitem
\bibitem[Kumar and Kumar(2010)Kumar, and Kumar]{kumar2010cathodes}
Kumar,~B.; Kumar,~J. Cathodes for solid-state lithium--oxygen cells: roles of
  NASICON glass-ceramics. \emph{Journal of the Electrochemical Society}
  \textbf{2010}, \emph{157}, A611\relax
\mciteBstWouldAddEndPuncttrue
\mciteSetBstMidEndSepPunct{\mcitedefaultmidpunct}
{\mcitedefaultendpunct}{\mcitedefaultseppunct}\relax
\EndOfBibitem
\bibitem[Kichambare \latin{et~al.}(2012)Kichambare, Rodrigues, and
  Kumar]{kichambare2012mesoporous}
Kichambare,~P.; Rodrigues,~S.; Kumar,~J. Mesoporous nitrogen-doped carbon-glass
  ceramic cathodes for solid-state lithium--oxygen batteries. \emph{ACS applied
  materials \& interfaces} \textbf{2012}, \emph{4}, 49--52\relax
\mciteBstWouldAddEndPuncttrue
\mciteSetBstMidEndSepPunct{\mcitedefaultmidpunct}
{\mcitedefaultendpunct}{\mcitedefaultseppunct}\relax
\EndOfBibitem
\bibitem[Kanno and Murayama(2001)Kanno, and Murayama]{kanno2001lithium}
Kanno,~R.; Murayama,~M. Lithium ionic conductor thio-LISICON: the Li2 S GeS2 P
  2 S 5 system. \emph{Journal of the electrochemical society} \textbf{2001},
  \emph{148}, A742\relax
\mciteBstWouldAddEndPuncttrue
\mciteSetBstMidEndSepPunct{\mcitedefaultmidpunct}
{\mcitedefaultendpunct}{\mcitedefaultseppunct}\relax
\EndOfBibitem
\bibitem[Kobayashi \latin{et~al.}(2008)Kobayashi, Yamada, and
  Kanno]{kobayashi2008interfacial}
Kobayashi,~T.; Yamada,~A.; Kanno,~R. Interfacial reactions at
  electrode/electrolyte boundary in all solid-state lithium battery using
  inorganic solid electrolyte, thio-LISICON. \emph{Electrochimica Acta}
  \textbf{2008}, \emph{53}, 5045--5050\relax
\mciteBstWouldAddEndPuncttrue
\mciteSetBstMidEndSepPunct{\mcitedefaultmidpunct}
{\mcitedefaultendpunct}{\mcitedefaultseppunct}\relax
\EndOfBibitem
\bibitem[Mizuno \latin{et~al.}(2005)Mizuno, Hayashi, Tadanaga, and
  Tatsumisago]{mizuno2005new}
Mizuno,~F.; Hayashi,~A.; Tadanaga,~K.; Tatsumisago,~M. New, highly
  ion-conductive crystals precipitated from Li $\{$sub 2$\}$ SP $\{$sub 2$\}$ S
  $\{$sub 5$\}$ glasses. \emph{Advanced Materials (Weinheim)} \textbf{2005},
  \emph{17}\relax
\mciteBstWouldAddEndPuncttrue
\mciteSetBstMidEndSepPunct{\mcitedefaultmidpunct}
{\mcitedefaultendpunct}{\mcitedefaultseppunct}\relax
\EndOfBibitem
\bibitem[Kondo \latin{et~al.}(1992)Kondo, Takada, and Yamamura]{kondo1992new}
Kondo,~S.; Takada,~K.; Yamamura,~Y. New lithium ion conductors based on
  Li2S-SiS2 system. \emph{Solid State Ionics} \textbf{1992}, \emph{53},
  1183--1186\relax
\mciteBstWouldAddEndPuncttrue
\mciteSetBstMidEndSepPunct{\mcitedefaultmidpunct}
{\mcitedefaultendpunct}{\mcitedefaultseppunct}\relax
\EndOfBibitem
\bibitem[Takada \latin{et~al.}(1993)Takada, Aotani, and
  Kondo]{takada1993electrochemical}
Takada,~K.; Aotani,~N.; Kondo,~S. Electrochemical behaviors of Li+ ion
  conductor, Li3PO4-Li2S-SiS2. \emph{Journal of power sources} \textbf{1993},
  \emph{43}, 135--141\relax
\mciteBstWouldAddEndPuncttrue
\mciteSetBstMidEndSepPunct{\mcitedefaultmidpunct}
{\mcitedefaultendpunct}{\mcitedefaultseppunct}\relax
\EndOfBibitem
\bibitem[Fergus(2010)]{fergus2010ceramic}
Fergus,~J.~W. Ceramic and polymeric solid electrolytes for lithium-ion
  batteries. \emph{Journal of Power Sources} \textbf{2010}, \emph{195},
  4554--4569\relax
\mciteBstWouldAddEndPuncttrue
\mciteSetBstMidEndSepPunct{\mcitedefaultmidpunct}
{\mcitedefaultendpunct}{\mcitedefaultseppunct}\relax
\EndOfBibitem
\bibitem[Cheng \latin{et~al.}(2018)Cheng, Liu, Mehta, Xin, Lin, Persson, Chen,
  Crumlin, and Doeff]{Cheng2018_garnet_LHX}
Cheng,~L.; Liu,~M.; Mehta,~A.; Xin,~H.; Lin,~F.; Persson,~K.; Chen,~G.;
  Crumlin,~E.~J.; Doeff,~M. {Garnet Electrolyte Surface Degradation and
  Recovery}. \emph{ACS Applied Energy Materials} \textbf{2018}, \emph{1},
  7244--7252\relax
\mciteBstWouldAddEndPuncttrue
\mciteSetBstMidEndSepPunct{\mcitedefaultmidpunct}
{\mcitedefaultendpunct}{\mcitedefaultseppunct}\relax
\EndOfBibitem
\bibitem[Wang \latin{et~al.}(2024)Wang, Barks, Lin, Xu, Melamed, McConohy,
  Nemsak, and Chueh]{Wang2024_garnet_LHX}
Wang,~S.; Barks,~E.; Lin,~P.~T.; Xu,~X.; Melamed,~C.; McConohy,~G.; Nemsak,~S.;
  Chueh,~W.~C. {Effect of H+ Exchange and Surface Impurities on Bulk and
  Interfacial Electrochemistry of Garnet Solid Electrolytes}. \emph{Chemistry
  of Materials} \textbf{2024}, \emph{36}, 6849--6864\relax
\mciteBstWouldAddEndPuncttrue
\mciteSetBstMidEndSepPunct{\mcitedefaultmidpunct}
{\mcitedefaultendpunct}{\mcitedefaultseppunct}\relax
\EndOfBibitem
\bibitem[Zhou \latin{et~al.}(2024)Zhou, Zhang, Zhang, and
  Liu]{Zhou2024_garnet_LHX}
Zhou,~J.; Zhang,~H.; Zhang,~L.; Liu,~Z. {Dynamical evolution of CO 2 and H 2 O
  on garnet electrolyte elucidated by ambient pressure X-ray spectroscopies}.
  \emph{Nature Communications} \textbf{2024}, \emph{15}, 2777\relax
\mciteBstWouldAddEndPuncttrue
\mciteSetBstMidEndSepPunct{\mcitedefaultmidpunct}
{\mcitedefaultendpunct}{\mcitedefaultseppunct}\relax
\EndOfBibitem
\bibitem[Zuo \latin{et~al.}(2022)Zuo, Xiao, Zarrabeitia, Xue, Yang, and
  Passerini]{zuo2022guidelines}
Zuo,~W.; Xiao,~Z.; Zarrabeitia,~M.; Xue,~X.; Yang,~Y.; Passerini,~S. Guidelines
  for air-stable lithium/sodium layered oxide cathodes. \emph{ACS Materials
  Letters} \textbf{2022}, \emph{4}, 1074--1086\relax
\mciteBstWouldAddEndPuncttrue
\mciteSetBstMidEndSepPunct{\mcitedefaultmidpunct}
{\mcitedefaultendpunct}{\mcitedefaultseppunct}\relax
\EndOfBibitem
\bibitem[Chen \latin{et~al.}(2019)Chen, Li, Yu, Chen, and
  Li]{chen2019approaching}
Chen,~R.; Li,~Q.; Yu,~X.; Chen,~L.; Li,~H. Approaching practically accessible
  solid-state batteries: stability issues related to solid electrolytes and
  interfaces. \emph{Chemical reviews} \textbf{2019}, \emph{120},
  6820--6877\relax
\mciteBstWouldAddEndPuncttrue
\mciteSetBstMidEndSepPunct{\mcitedefaultmidpunct}
{\mcitedefaultendpunct}{\mcitedefaultseppunct}\relax
\EndOfBibitem
\bibitem[Shimonishi \latin{et~al.}(2011)Shimonishi, Zhang, Imanishi, Im, Lee,
  Hirano, Takeda, Yamamoto, and Sammes]{Shimonishi2011_NASICON_alkaline}
Shimonishi,~Y.; Zhang,~T.; Imanishi,~N.; Im,~D.; Lee,~D.~J.; Hirano,~A.;
  Takeda,~Y.; Yamamoto,~O.; Sammes,~N. {A study on lithium/air secondary
  batteries - Stability of the NASICON-type lithium ion conducting solid
  electrolyte in alkaline aqueous solutions}. \emph{Journal of Power Sources}
  \textbf{2011}, \emph{196}, 5128--5132\relax
\mciteBstWouldAddEndPuncttrue
\mciteSetBstMidEndSepPunct{\mcitedefaultmidpunct}
{\mcitedefaultendpunct}{\mcitedefaultseppunct}\relax
\EndOfBibitem
\bibitem[Zhang \latin{et~al.}(2013)Zhang, Matsui, Hirano, Takeda, Yamamoto, and
  Imanishi]{Zhang2013_NASICON_water}
Zhang,~P.; Matsui,~M.; Hirano,~A.; Takeda,~Y.; Yamamoto,~O.; Imanishi,~N.
  {Water-stable lithium ion conducting solid electrolyte of the NASICON-type
  structure}. \emph{Solid State Ionics} \textbf{2013}, \emph{253},
  175--180\relax
\mciteBstWouldAddEndPuncttrue
\mciteSetBstMidEndSepPunct{\mcitedefaultmidpunct}
{\mcitedefaultendpunct}{\mcitedefaultseppunct}\relax
\EndOfBibitem
\bibitem[He \latin{et~al.}(2014)He, Zu, Wang, Han, Yin, Zhao, Liu, and
  Chen]{He2014_LAGP_water}
He,~K.; Zu,~C.; Wang,~Y.; Han,~B.; Yin,~X.; Zhao,~H.; Liu,~Y.; Chen,~J.
  {Stability of lithium ion conductor NASICON structure glass ceramic in acid
  and alkaline aqueous solution}. \emph{Solid State Ionics} \textbf{2014},
  \emph{254}, 78--81\relax
\mciteBstWouldAddEndPuncttrue
\mciteSetBstMidEndSepPunct{\mcitedefaultmidpunct}
{\mcitedefaultendpunct}{\mcitedefaultseppunct}\relax
\EndOfBibitem
\bibitem[Safanama and Adams(2017)Safanama, and Adams]{Safanama2017_LAGP_water}
Safanama,~D.; Adams,~S. {High efficiency aqueous and hybrid lithium-air
  batteries enabled by Li1.5Al0.5Ge1.5(PO4)3 ceramic anode-protecting
  membranes}. \emph{Journal of Power Sources} \textbf{2017}, \emph{340},
  294--301\relax
\mciteBstWouldAddEndPuncttrue
\mciteSetBstMidEndSepPunct{\mcitedefaultmidpunct}
{\mcitedefaultendpunct}{\mcitedefaultseppunct}\relax
\EndOfBibitem
\bibitem[Kim \latin{et~al.}(2023)Kim, Lee, Choi, Yoon, Jung, Kim, Kim, Park,
  and Im]{kim2023operando}
Kim,~H.; Lee,~H.; Choi,~W.; Yoon,~G.; Jung,~C.; Kim,~M.; Kim,~T.; Park,~J.;
  Im,~D. Operando Observation of the De-Evolution/Evolution Process of Hydrated
  LiOH in Moisture-Assisted Li--O2 Batteries. \emph{ACS Applied Materials \&
  Interfaces} \textbf{2023}, \emph{15}, 29120--29126\relax
\mciteBstWouldAddEndPuncttrue
\mciteSetBstMidEndSepPunct{\mcitedefaultmidpunct}
{\mcitedefaultendpunct}{\mcitedefaultseppunct}\relax
\EndOfBibitem
\bibitem[Monnin and Dubois(2005)Monnin, and Dubois]{monnin2005thermodynamics}
Monnin,~C.; Dubois,~M. Thermodynamics of the LiOH+ H2O system. \emph{Journal of
  Chemical \& Engineering Data} \textbf{2005}, \emph{50}, 1109--1113\relax
\mciteBstWouldAddEndPuncttrue
\mciteSetBstMidEndSepPunct{\mcitedefaultmidpunct}
{\mcitedefaultendpunct}{\mcitedefaultseppunct}\relax
\EndOfBibitem
\bibitem[Hasegawa \latin{et~al.}(2009)Hasegawa, Imanishi, Zhang, Xie, Hirano,
  Takeda, and Yamamoto]{hasegawa2009study}
Hasegawa,~S.; Imanishi,~N.; Zhang,~T.; Xie,~J.; Hirano,~A.; Takeda,~Y.;
  Yamamoto,~O. Study on lithium/air secondary batteries—Stability of
  NASICON-type lithium ion conducting glass--ceramics with water. \emph{Journal
  of power sources} \textbf{2009}, \emph{189}, 371--377\relax
\mciteBstWouldAddEndPuncttrue
\mciteSetBstMidEndSepPunct{\mcitedefaultmidpunct}
{\mcitedefaultendpunct}{\mcitedefaultseppunct}\relax
\EndOfBibitem
\bibitem[Dermenci(2020)]{dermenci2020stability}
Dermenci,~K.~B. Stability of Solid-State Sintered Li1. 5Al0. 5Ge1. 5 (PO4) 3
  Solid Electrolytes in Var-ious Mediums for All Solid-State Li-ion Batteries.
  \emph{International Journal of Automotive Science and Technology}
  \textbf{2020}, \emph{4}, 295--299\relax
\mciteBstWouldAddEndPuncttrue
\mciteSetBstMidEndSepPunct{\mcitedefaultmidpunct}
{\mcitedefaultendpunct}{\mcitedefaultseppunct}\relax
\EndOfBibitem
\bibitem[Lam \latin{et~al.}(2024)Lam, Li, Mishra, and
  Ceder]{Lam2024_ltgp_degradation}
Lam,~B.~X.; Li,~Z.; Mishra,~T.~P.; Ceder,~G. {Degradation Mechanism of
  Phosphate-Based Li-NASICON Conductors in Alkaline Environment}.
  \emph{Advanced Energy Materials} \textbf{2024}, \emph{2403596}\relax
\mciteBstWouldAddEndPuncttrue
\mciteSetBstMidEndSepPunct{\mcitedefaultmidpunct}
{\mcitedefaultendpunct}{\mcitedefaultseppunct}\relax
\EndOfBibitem
\bibitem[Hou \latin{et~al.}(2021)Hou, Yang, Liang, Dong, Chen, Li, Chen, Dai,
  and Xue]{hou2021multiscale}
Hou,~M.; Yang,~X.; Liang,~F.; Dong,~P.; Chen,~Y.; Li,~J.; Chen,~K.; Dai,~Y.;
  Xue,~D. Multiscale investigation into chemically stable NASICON solid
  electrolyte in acidic solutions. \emph{ACS Applied Materials \& Interfaces}
  \textbf{2021}, \emph{13}, 33262--33271\relax
\mciteBstWouldAddEndPuncttrue
\mciteSetBstMidEndSepPunct{\mcitedefaultmidpunct}
{\mcitedefaultendpunct}{\mcitedefaultseppunct}\relax
\EndOfBibitem
\bibitem[Abdullahi \latin{et~al.}(2024)Abdullahi, Tseng, Modak, Ismail,
  Sakamoto, and Kwabi]{abdullahi2024nasicon}
Abdullahi,~I.~M.; Tseng,~K.-T.; Modak,~S.~V.; Ismail,~Z.; Sakamoto,~J.;
  Kwabi,~D.~G. NaSICON Prepared by a Solution-Assisted Reaction Shows Enhanced
  Stability for High-Voltage Aqueous Redox-Flow Batteries. \emph{ACS Applied
  Energy Materials} \textbf{2024}, \emph{7}, 11724--11732\relax
\mciteBstWouldAddEndPuncttrue
\mciteSetBstMidEndSepPunct{\mcitedefaultmidpunct}
{\mcitedefaultendpunct}{\mcitedefaultseppunct}\relax
\EndOfBibitem
\bibitem[Fryer and Lad(2016)Fryer, and Lad]{fryer2016synthesis}
Fryer,~R.~T.; Lad,~R.~J. Synthesis and thermal stability of Pt3Si, Pt2Si, and
  PtSi films grown by e-beam co-evaporation. \emph{Journal of Alloys and
  Compounds} \textbf{2016}, \emph{682}, 216--224\relax
\mciteBstWouldAddEndPuncttrue
\mciteSetBstMidEndSepPunct{\mcitedefaultmidpunct}
{\mcitedefaultendpunct}{\mcitedefaultseppunct}\relax
\EndOfBibitem
\bibitem[Choi \latin{et~al.}(2024)Choi, Park, Engelhard, Li, Sushko, and
  Du]{choi2024reevaluation}
Choi,~M.-J.; Park,~H.; Engelhard,~M.~H.; Li,~D.; Sushko,~P.~V.; Du,~Y.
  Reevaluation of XPS Pt 4f peak fitting: Ti 3s plasmon peak interference and
  Pt metallic peak asymmetry in Pt@ TiO2 system. \emph{Journal of Vacuum
  Science \& Technology A} \textbf{2024}, \emph{42}\relax
\mciteBstWouldAddEndPuncttrue
\mciteSetBstMidEndSepPunct{\mcitedefaultmidpunct}
{\mcitedefaultendpunct}{\mcitedefaultseppunct}\relax
\EndOfBibitem
\bibitem[Yao \latin{et~al.}(2013)Yao, Kwabi, Quinlan, Mansour, Grimaud, Lee,
  Lu, and Shao-Horn]{yao2013thermal}
Yao,~K.~P.; Kwabi,~D.~G.; Quinlan,~R.~A.; Mansour,~A.~N.; Grimaud,~A.;
  Lee,~Y.-L.; Lu,~Y.-C.; Shao-Horn,~Y. Thermal stability of Li2O2 and Li2O for
  Li-air batteries: in situ XRD and XPS studies. \emph{Journal of The
  Electrochemical Society} \textbf{2013}, \emph{160}, A824\relax
\mciteBstWouldAddEndPuncttrue
\mciteSetBstMidEndSepPunct{\mcitedefaultmidpunct}
{\mcitedefaultendpunct}{\mcitedefaultseppunct}\relax
\EndOfBibitem
\bibitem[Lindblad \latin{et~al.}(1994)Lindblad, Rebenstorf, Yan, and
  Andersson]{lindblad1994characterization}
Lindblad,~T.; Rebenstorf,~B.; Yan,~Z.-G.; Andersson,~S. L.~T. Characterization
  of vanadia supported on amorphous AlPO4 and its properties for oxidative
  dehydrogenation of propane. \emph{Applied Catalysis A: General}
  \textbf{1994}, \emph{112}, 187--208\relax
\mciteBstWouldAddEndPuncttrue
\mciteSetBstMidEndSepPunct{\mcitedefaultmidpunct}
{\mcitedefaultendpunct}{\mcitedefaultseppunct}\relax
\EndOfBibitem
\bibitem[Morgan \latin{et~al.}(1973)Morgan, Van~Wazer, and
  Stec]{morgan1973inner}
Morgan,~W.~E.; Van~Wazer,~J.~R.; Stec,~W.~J. Inner-orbital photoelectron
  spectroscopy of the alkali metal halides, perchlorates, phosphates, and
  pyrophosphates. \emph{Journal of the American Chemical Society}
  \textbf{1973}, \emph{95}, 751--755\relax
\mciteBstWouldAddEndPuncttrue
\mciteSetBstMidEndSepPunct{\mcitedefaultmidpunct}
{\mcitedefaultendpunct}{\mcitedefaultseppunct}\relax
\EndOfBibitem
\bibitem[Muto \latin{et~al.}(2021)Muto, Yamamoto, Sakakura, Tian, Tateyama, and
  Iriyama]{muto2021stem}
Muto,~S.; Yamamoto,~Y.; Sakakura,~M.; Tian,~H.-K.; Tateyama,~Y.; Iriyama,~Y.
  STEM-EELS spectrum imaging of an aerosol-deposited NASICON-type LATP solid
  electrolyte and LCO cathode interface. \emph{ACS Applied Energy Materials}
  \textbf{2021}, \emph{5}, 98--107\relax
\mciteBstWouldAddEndPuncttrue
\mciteSetBstMidEndSepPunct{\mcitedefaultmidpunct}
{\mcitedefaultendpunct}{\mcitedefaultseppunct}\relax
\EndOfBibitem
\bibitem[Gao \latin{et~al.}(2014)Gao, Gu, Nie, Mashayek, Wang, Odegard, and
  Shahbazian-Yassar]{gao2014direct}
Gao,~Q.; Gu,~M.; Nie,~A.; Mashayek,~F.; Wang,~C.; Odegard,~G.~M.;
  Shahbazian-Yassar,~R. Direct evidence of lithium-induced atomic ordering in
  amorphous TiO2 nanotubes. \emph{Chemistry of Materials} \textbf{2014},
  \emph{26}, 1660--1669\relax
\mciteBstWouldAddEndPuncttrue
\mciteSetBstMidEndSepPunct{\mcitedefaultmidpunct}
{\mcitedefaultendpunct}{\mcitedefaultseppunct}\relax
\EndOfBibitem
\bibitem[Persson \latin{et~al.}(2012)Persson, Waldwick, Lazic, and
  Ceder]{Persson2012_pbx}
Persson,~K.~A.; Waldwick,~B.; Lazic,~P.; Ceder,~G. {Prediction of solid-aqueous
  equilibria : Scheme to combine first-principles calculations of solids with
  experimental aqueous states}. \emph{Physical Review B} \textbf{2012},
  \emph{85}, 235438\relax
\mciteBstWouldAddEndPuncttrue
\mciteSetBstMidEndSepPunct{\mcitedefaultmidpunct}
{\mcitedefaultendpunct}{\mcitedefaultseppunct}\relax
\EndOfBibitem
\bibitem[Singh \latin{et~al.}(2017)Singh, Zhou, Shinde, Suram, Montoya,
  Winston, Gregoire, and Persson]{Singh2017_pbx}
Singh,~A.~K.; Zhou,~L.; Shinde,~A.; Suram,~S.~K.; Montoya,~J.~H.; Winston,~D.;
  Gregoire,~J.~M.; Persson,~K.~A. {Electrochemical Stability of Metastable
  Materials}. \emph{Chemistry of Materials} \textbf{2017}, \emph{29},
  10159−10167\relax
\mciteBstWouldAddEndPuncttrue
\mciteSetBstMidEndSepPunct{\mcitedefaultmidpunct}
{\mcitedefaultendpunct}{\mcitedefaultseppunct}\relax
\EndOfBibitem
\bibitem[Sun \latin{et~al.}(2019)Sun, Kitchaev, Kramer, and Ceder]{Sun2019_pbx}
Sun,~W.; Kitchaev,~D.~A.; Kramer,~D.; Ceder,~G. {Non-equilibrium
  crystallization pathways of manganese oxides in aqueous solution}.
  \emph{Nature Communications} \textbf{2019}, \emph{10}, 573\relax
\mciteBstWouldAddEndPuncttrue
\mciteSetBstMidEndSepPunct{\mcitedefaultmidpunct}
{\mcitedefaultendpunct}{\mcitedefaultseppunct}\relax
\EndOfBibitem
\bibitem[Tian \latin{et~al.}(2019)Tian, Liu, Ji, Chen, and
  Qi]{Tian2019_latp_surface}
Tian,~H.~K.; Liu,~Z.; Ji,~Y.; Chen,~L.~Q.; Qi,~Y. {Interfacial Electronic
  Properties Dictate Li Dendrite Growth in Solid Electrolytes}. \emph{Chemistry
  of Materials} \textbf{2019}, \emph{31}, 7351--7359\relax
\mciteBstWouldAddEndPuncttrue
\mciteSetBstMidEndSepPunct{\mcitedefaultmidpunct}
{\mcitedefaultendpunct}{\mcitedefaultseppunct}\relax
\EndOfBibitem
\bibitem[Pogosova \latin{et~al.}(2020)Pogosova, Krasnikova, Sanin, Lipovskikh,
  Eliseev, Sergeev, and Stevenson]{Pogosova2020_latp_surface}
Pogosova,~M.~A.; Krasnikova,~I.~V.; Sanin,~A.~O.; Lipovskikh,~S.~A.;
  Eliseev,~A.~A.; Sergeev,~A.~V.; Stevenson,~K.~J. {Complex Investigation of
  Water Impact on Li-Ion Conductivity of Li1.3Al0.3Ti1.7(PO4)3-
  Electrochemical, Chemical, Structural, and Morphological Aspects}.
  \emph{Chemistry of Materials} \textbf{2020}, \emph{32}, 3723--3732\relax
\mciteBstWouldAddEndPuncttrue
\mciteSetBstMidEndSepPunct{\mcitedefaultmidpunct}
{\mcitedefaultendpunct}{\mcitedefaultseppunct}\relax
\EndOfBibitem
\bibitem[Zhu \latin{et~al.}(2015)Zhu, Zhao, Wei, Tan, and An]{zhu2015high}
Zhu,~X.; Zhao,~T.; Wei,~Z.; Tan,~P.; An,~L. A high-rate and long cycle life
  solid-state lithium--air battery. \emph{Energy \& Environmental Science}
  \textbf{2015}, \emph{8}, 3745--3754\relax
\mciteBstWouldAddEndPuncttrue
\mciteSetBstMidEndSepPunct{\mcitedefaultmidpunct}
{\mcitedefaultendpunct}{\mcitedefaultseppunct}\relax
\EndOfBibitem
\bibitem[Tio \latin{et~al.}(2019)Tio, Ulkuniemi, Nyysso, Lahtonen, and
  Valden]{Tio2019_TiO2}
Tio,~A.-l.-d.~B.; Ulkuniemi,~R.; Nyysso,~T.; Lahtonen,~K.; Valden,~M.
  {Diversity of TiO 2 : Controlling the Molecular and Electronic Structure of
  Atomic-Layer-Deposited Black TiO 2}. \emph{ACS Applied Materials \&
  Interfaces} \textbf{2019}, \emph{11}, 2758--2762\relax
\mciteBstWouldAddEndPuncttrue
\mciteSetBstMidEndSepPunct{\mcitedefaultmidpunct}
{\mcitedefaultendpunct}{\mcitedefaultseppunct}\relax
\EndOfBibitem
\bibitem[Acevedo-pe{\~{n}}a \latin{et~al.}(2013)Acevedo-pe{\~{n}}a,
  Electrochem, Soc, Vazquez-arenas, Lartundo-rojas, and
  Gonz]{Acevedo-pena2013_TiO2}
Acevedo-pe{\~{n}}a,~P.; Electrochem,~J.; Soc,~C.; Vazquez-arenas,~J.;
  Lartundo-rojas,~L.; Gonz,~I. {Ti Anodization in Alkaline Electrolyte : The
  Relationship between Transport of Defects , Film Hydration and Composition Ti
  Anodization in Alkaline Electrolyte : The Relationship between Transport of
  Defects , Film Hydration and Composition}. \emph{Journal of The
  Electrochemical Society} \textbf{2013}, \emph{160}, C277--C284\relax
\mciteBstWouldAddEndPuncttrue
\mciteSetBstMidEndSepPunct{\mcitedefaultmidpunct}
{\mcitedefaultendpunct}{\mcitedefaultseppunct}\relax
\EndOfBibitem
\bibitem[Arunachalam \latin{et~al.}(2023)Arunachalam, Kanase, Zhu, and
  Kang]{arunachalam2023reliable}
Arunachalam,~M.; Kanase,~R.~S.; Zhu,~K.; Kang,~S.~H. Reliable bi-functional
  nickel-phosphate/TiO2 integration enables stable n-GaAs photoanode for water
  oxidation under alkaline condition. \emph{Nature Communications}
  \textbf{2023}, \emph{14}, 5429\relax
\mciteBstWouldAddEndPuncttrue
\mciteSetBstMidEndSepPunct{\mcitedefaultmidpunct}
{\mcitedefaultendpunct}{\mcitedefaultseppunct}\relax
\EndOfBibitem
\bibitem[Wood and Teeter(2018)Wood, and Teeter]{wood2018xps}
Wood,~K.~N.; Teeter,~G. XPS on Li-battery-related compounds: analysis of
  inorganic SEI phases and a methodology for charge correction. \emph{ACS
  Applied Energy Materials} \textbf{2018}, \emph{1}, 4493--4504\relax
\mciteBstWouldAddEndPuncttrue
\mciteSetBstMidEndSepPunct{\mcitedefaultmidpunct}
{\mcitedefaultendpunct}{\mcitedefaultseppunct}\relax
\EndOfBibitem
\bibitem[Kre(1996)]{Kresse1996_vasp}
{Efficient iterative schemes for ab initio total-energy calculations using a
  plane-wave basis set}. \emph{Physical Review B - Condensed Matter and
  Materials Physics} \textbf{1996}, \emph{54}, 11169--11186\relax
\mciteBstWouldAddEndPuncttrue
\mciteSetBstMidEndSepPunct{\mcitedefaultmidpunct}
{\mcitedefaultendpunct}{\mcitedefaultseppunct}\relax
\EndOfBibitem
\bibitem[Kresse and Joubert(1999)Kresse, and Joubert]{Kresse1996_paw}
Kresse,~G.; Joubert,~D. {From ultrasoft pseudopotentials to the projector
  augmented-wave method}. \emph{Phys. Rev. B} \textbf{1999}, \emph{59},
  1758--1775\relax
\mciteBstWouldAddEndPuncttrue
\mciteSetBstMidEndSepPunct{\mcitedefaultmidpunct}
{\mcitedefaultendpunct}{\mcitedefaultseppunct}\relax
\EndOfBibitem
\bibitem[Bart{\'{o}}k and Yates(2019)Bart{\'{o}}k, and
  Yates]{Bartok2019_r2scan}
Bart{\'{o}}k,~A.~P.; Yates,~J.~R. {Regularized SCAN functional}. \emph{Journal
  of Chemical Physics} \textbf{2019}, \emph{150}\relax
\mciteBstWouldAddEndPuncttrue
\mciteSetBstMidEndSepPunct{\mcitedefaultmidpunct}
{\mcitedefaultendpunct}{\mcitedefaultseppunct}\relax
\EndOfBibitem
\bibitem[Perdew \latin{et~al.}(1996)Perdew, Burke, and
  Ernzerhof]{Perdew1996_pbe}
Perdew,~J.~P.; Burke,~K.; Ernzerhof,~M. {Generalized gradient approximation
  made simple}. \emph{Physical Review Letters} \textbf{1996}, \emph{77},
  3865--3868\relax
\mciteBstWouldAddEndPuncttrue
\mciteSetBstMidEndSepPunct{\mcitedefaultmidpunct}
{\mcitedefaultendpunct}{\mcitedefaultseppunct}\relax
\EndOfBibitem
\bibitem[Ong \latin{et~al.}(2013)Ong, Richards, Jain, Hautier, Kocher, Cholia,
  Gunter, Chevrier, Persson, and Ceder]{Ong2013_pymatgen}
Ong,~S.~P.; Richards,~W.~D.; Jain,~A.; Hautier,~G.; Kocher,~M.; Cholia,~S.;
  Gunter,~D.; Chevrier,~V.~L.; Persson,~K.~A.; Ceder,~G. {Python Materials
  Genomics (pymatgen): A robust, open-source python library for materials
  analysis}. \emph{Computational Materials Science} \textbf{2013}, \emph{68},
  314--319\relax
\mciteBstWouldAddEndPuncttrue
\mciteSetBstMidEndSepPunct{\mcitedefaultmidpunct}
{\mcitedefaultendpunct}{\mcitedefaultseppunct}\relax
\EndOfBibitem
\end{mcitethebibliography}

\end{document}